\documentclass[fleqn,usenatbib]{mnras}

\usepackage[utf8]{inputenc}
\usepackage{enumitem}
\usepackage{graphicx}
\usepackage{hyperref}
\usepackage{mathtools}
\usepackage{comment}
\usepackage{amsmath}
\usepackage{ amssymb }
\usepackage{lipsum} 
\usepackage{float}
\usepackage{cleveref}
\usepackage[caption=false]{subfig}
\usepackage{placeins}
\usepackage{xcolor}
\usepackage{booktabs}

\usepackage{comment}

\title[Probing feedback with the SZ]
{Inferring the impact of feedback on the matter distribution using the Sunyaev Zel'dovich effect: Insights from CAMELS simulations and ACT+DES data}

\author[Pandey et al.]{Shivam Pandey,$^{1,2}$
Kai Lehman,$^{3,4}$
Eric J. Baxter,$^{3}$
Yueying Ni,$^{5}$
\newauthor
Daniel Angl{\'e}s-Alc{\'a}zar,$^{6,7}$
Shy Genel,$^{7,1}$
Francisco Villaescusa-Navarro,$^{7}$
\newauthor
Ana Maria Delgado,$^{5}$
Tiziana di Matteo$^{9}$
\\$^{1}$Department of Physics, Columbia University, New York, NY, USA 10027
\\$^{2}$Department of Physics and Astronomy, University of Pennsylvania, Philadelphia, PA 19104, USA
\\$^{3}$Institute for Astronomy, University of Hawai`i, 2680 Woodlawn Drive, Honolulu, HI 96822, USA
\\$^{4}$Universitäts-Sternwarte München, Fakultät für Physik, Ludwig-Maximilians-Universität, Scheinerstr. 1, 81679 München, Germany
\\$^{5}$Center for Astrophysics {$|$} Harvard \& Smithsonian, Cambridge, MA 02138, US
\\$^{6}$Department of Physics, University of Connecticut, 196 Auditorium Road, U-3046, Storrs, CT, 06269, USA
\\$^{7}$Center for Computational Astrophysics, Flatiron Institute, 162 5th Avenue, New York, NY, 10010, USA
\\$^{9 }$McWilliams Center for Cosmology, Department of Physics, Carnegie Mellon University, Pittsburgh, PA 15213
}

\begin{document}

\label{firstpage}

\pagerange{\pageref{firstpage}--\pageref{lastpage}}
\maketitle

\begin{abstract}
Feedback from active galactic nuclei and stellar processes changes the matter distribution on small scales, leading to significant systematic uncertainty in weak lensing constraints on cosmology.  We investigate how the observable properties of group-scale halos can constrain feedback's impact on the matter distribution using  Cosmology and Astrophysics with MachinE Learning Simulations (CAMELS).  Extending the results of previous work to smaller halo masses and higher wavenumber, $k$, we find that the baryon fraction in halos contains significant information about the impact of feedback on the matter power spectrum.  We explore how the thermal Sunyaev Zel'dovich (tSZ) signal from group-scale halos contains similar information. Using recent Dark Energy Survey (DES) weak lensing and Atacama Cosmology Telescope (ACT) tSZ cross-correlation measurements and models trained on CAMELS, we obtain 10\% constraints on feedback effects on the power spectrum at $k \sim 5\, h/{\rm Mpc}$. We show that with future surveys, it will be possible to constrain baryonic effects on the power spectrum to $\mathcal{O}(<1\%)$ at $k = 1\, h/{\rm Mpc}$ and $\mathcal{O}(3\%)$ at $k = 5\, h/{\rm Mpc}$ using the methods that we introduce here.
Finally, we investigate the impact of feedback on the matter bispectrum, finding that tSZ observables are highly informative in this case. 
\end{abstract}  


\section{Introduction}\label{sec:intro}

The statistics of the matter distribution on scales $k \gtrsim 0.1\,h{\rm Mpc}^{-1}$ are tightly constrained by current  weak lensing surveys \citep[e.g.][]{Asgari:2021,DESY3cosmo}.  However, modeling the matter distribution on these scales to extract cosmological information is complicated by the effects of baryonic feedback \citep{Rudd:2008}.  Energetic output from active galactic nuclei (AGN) and stellar processes (e.g. winds and supernovae) directly impacts the distribution of gas on small scales, thereby changing the total matter distribution \citep[e.g.][]{Chisari:2019}.\footnote{Changes to the gas distribution can also gravitationally influence the dark matter distribution, further modifying the total matter distribution.}  
The coupling between these processes and the large-scale gas distribution is challenging to model theoretically and in simulations because of the large dynamic range involved, from the scales of individual stars to the scales of galaxy clusters.  While it is generally agreed that feedback leads to a suppression of the matter power spectrum on scales $0.1\,h{\rm Mpc}^{-1} \lesssim k \lesssim 20\,h{\rm Mpc}^{-1}$, the amplitude of this suppression remains uncertain by tens of percent \citep{vanDaalen:2020, Villaescusa-Navarro:2021:ApJ:} (see also Fig.~\ref{fig:Pk_Bk_CV}). This systematic uncertainty  limits constraints on cosmological parameters from current weak lensing surveys \cite[e.g.][]{DESY3cosmo,Asgari:2021}.  For future surveys, such as the Vera Rubin Observatory LSST \citep{TheLSSTDarkEnergyScienceCollaboration:2018:arXiv:} and \textit{Euclid} \citep{EuclidCollaboration:2020:A&A:}, the problem will become even more severe given  expected increases in statistical precision.  In order to reduce the systematic uncertainties associated with feedback, we would like to identify observable quantities that carry information about the impact of feedback on the matter distribution and develop approaches to extract this information \citep[e.g.][]{Nicola:2022:JCAP:}. 

Recently, \citet{vanDaalen:2020} showed that the halo baryon fraction, $f_b$, in halos with $M \sim 10^{14}\,M_{\odot}$ carries significant information about suppression of the matter power spectrum caused by baryonic feedback. They found that the relation between $f_b$ and matter power suppression was robust to at least some changes in the subgrid prescriptions for feedback physics. Note that $f_b$ as defined by \citet{vanDaalen:2020} counts baryons in both the intracluster medium as well as those in stars. The connection between $f_b$ and feedback is expected, since one of the main drivers of feedback's impact on the matter distribution is the ejection of gas from halos by AGN.  Therefore, when feedback is strong, halos will be depleted of baryons and $f_b$ will be lower.  The conversion of baryons into stars --- which will not significantly impact the matter power spectrum on large scales --- does not impact $f_b$, since $f_b$ includes baryons in stars as well as the ICM. 
\citet{vanDaalen:2020} specifically consider the measurement of $f_b$ in halos with $6\times 10^{13} M_{\odot} \lesssim M_{500c} \lesssim 10^{14}\,M_{\odot}$.  In much more massive halos, the energy output of AGN is small compared to the binding energy of the halo, preventing gas from being expelled.  In smaller halos, \citet{vanDaalen:2020} found that the correlation between power spectrum suppression and $f_b$ is less clear. 

Although $f_b$ carries information about feedback, it is somewhat unclear how one would measure $f_b$ in practice.  Observables such as the kinematic Sunyaev Zel'dovich (kSZ) effect can be used to constrain the gas density; combined with some estimate of stellar mass, $f_b$ could then be inferred.  However, measuring the kSZ is challenging, and current measurements have low signal-to-noise \citep{Hand:2012,Hill:2016,Soergel:2016}.  Moreover, \citet{vanDaalen:2020} consider a relatively limited range of feedback prescriptions.  It is unclear whether a broader range of feedback models could lead to a greater spread in the relationship between $f_b$ and baryonic effects on the power spectrum.  In any case, it is worthwhile to consider other potential observational probes of feedback.

Another potentially powerful probe of baryonic feedback is the thermal SZ (tSZ) effect.  The tSZ effect is caused by inverse Compton scattering of CMB photons with a population of electrons at high temperature.  This scattering process leads to a spectral distortion in the CMB that can be reconstructed from multi-frequency CMB observations.  The amplitude of this distortion is sensitive to the line-of-sight integral of the electron pressure.  Since feedback changes the distribution and thermodynamics of the gas, it stands to reason that it could impact the tSZ signal.   Indeed, several works using both data \citep[e.g][]{Pandey:2019,Pandey:2022,Gatti:2022} and simulations \citep[e.g.][]{Scannapieco:2008,Bhattacharya:2008,Moser:2022,Wadekar:2022} have shown that the tSZ signal from low-mass (group scale) halos is sensitive to feedback.  Excitingly, the sensitivity of tSZ measurements is expected to increase dramatically in the near future due to high-sensitivity CMB measurements from e.g. SPT-3G \citep{Benson:2014:SPIE:}, Advanced ACTPol \citep{Henderson:2016:JLTP:},  Simons Observatory \citep{Ade:2019:JCAP:}, and CMB Stage 4 \citep{CMBS4}.

The goal of this work is to investigate what information the tSZ signals from low-mass halos contain about the impact  of feedback on the small-scale matter distribution.  The tSZ signal, which we denote with the Compton $y$ parameter, carries different information from $f_b$.  For one, $y$  is sensitive only to the gas and not to stellar mass.  Moreover, $y$ carries sensitivity to both the gas density and temperature, unlike $f_b$ which depends only on the gas density.  The $y$ signal is also easier to measure than $f_b$, since it can be estimated simply by cross-correlating halos with a tSZ map.  The signal-to-noise of such cross-correlation measurements is already high with current data, on the order of 10s of $\sigma$ \citep{Vikram:2017,Pandey:2019,Pandey:2022,Sanchez:2022}.  

In this paper, we investigate the information content of the tSZ signal from group-scale halos using the Cosmology and Astrophysics with MachinE Learning Simulations (CAMELS) simulations. As we describe in more detail in \S\ref{sec:camels}, CAMELS is a suite of many hydrodynamical simulations run across a range of different feedback prescriptions and different cosmological parameters.  The relatively small volume of the CAMELS simulations ($(25/h)^3\,{\rm Mpc^3}$) means that we are somewhat limited in the halo masses and scales that we can probe.  We therefore view our analysis as an exploratory work that investigates the information content of low-mass halos for constraining feedback and how to extract this information; more accurate results over a wider range of halo mass and $k$ may be obtained in the future using the same methods applied to larger volume simulations.

By training statistical models on the CAMELS simulations, we explore what information about feedback exists in tSZ observables, and how robust this information is to changes in subgrid feedback prescriptions.  We consider three very different prescriptions for feedback based on the SIMBA \citep{Dave:2019:MNRAS:}, Illustris-TNG (\citealt{Pillepich:2018:MNRAS:}, henceforth TNG) and Astrid \citep{Bird:2022:MNRAS:, Ni:2022:MNRAS:} models across a wide range of possible parameter values, including variations in cosmology.
The flexibility of the statistical models we employ means that it is possible to uncover more complex relationships between e.g. $f_b$, $y$, and the baryonic suppression of the power spectrum than considered in \citet{vanDaalen:2020}. The work presented here is complementary to \citet{Delgado:23} which explores the information content in the baryon fraction of halos encompassing broader mass range ($M > 10^{10} M_{\odot}/h$), finding a broad correlation with the matter power suppression. 

Finally, we apply our trained statistical models to recent measurements of the $y$ signal from low-mass halos by \citet{Gatti:2022} and \citet{Pandey:2022}.  These analyses inferred the halo-integrated $y$ signal from the cross-correlation of galaxy lensing and the tSZ effect using lensing data from the Dark Energy Survey (DES) \citep{Amon:2022:PhRvD:, Secco:2022:PhRvD:b} and tSZ measurements from the Atacama Cosmology Telescope (ACT) \citep{Madhavacheril:2020:PhRvD:}.  In addition to providing interesting constraints on the impact of feedback, these results highlight the potential of future similar analyses with e.g. Dark Energy Spectroscopic Experiment (DESI; \citealt{DESI}), Simons Observatory \citep{Ade:2019:JCAP:}, and CMB Stage 4 \citep{CMBS4}.

Two recent works --- \citet{Moser:2022} and \citet{Wadekar:2022} --- have used the CAMELS simulations to explore the information content of the tSZ signal for constraining feedback.  These works focus on the ability of tSZ observations to constrain the parameters of subgrid feedback models in hydrodynamical simulations.  Here, in contrast, we attempt to connect the observable quantities directly to the impact of feedback on the matter power spectrum and bispectrum.   Additionally, unlike some of the results presented in \citet{Moser:2022} and \citet{Wadekar:2022}, we consider the full parameter space explored by the CAMELS simulations rather than the small variations around a fiducial point that are relevant to calculation of the Fisher matrix. Finally, we only focus on the intra-halo gas profile of the halos in the mass range captured by the CAMELS simulations (c.f. \citealt{Moser:2022}). We do not expect the inter-halo gas pressure to be captured by the small boxes used here as it may be sensitive to higher halo masses \citep{Pandey:2020}. 

Nonlinear evolution of the matter distribution induces non-Gaussianity, and hence there is additional information to be recovered beyond the power spectrum. Recent measurements detect higher-order matter correlations at cosmological scales at $\mathcal{O}(10\sigma)$\citep{Secco:2022:PhRvD:b, Gatti:2022:PhRvD:}, and the significance of these measurements is expected to rapidly increase with up-coming surveys \citep{Pyne:2021:MNRAS:}. Jointly analyzing two-point and three-point correlations of the matter field can help with self-calibration of systematic parameters and improve cosmological constraints. As described in \citet{Foreman:2020:MNRAS:}, the matter bispectrum is expected to be impacted by baryonic physics at $\mathcal{O}(10\%)$ over the scales of interest.  With these considerations in mind, we also investigate whether the SZ observations carry information about the impact of baryonic feedback on the matter bispectrum. 

The plan of the paper is as follows.  In \S\ref{sec:camels} we discuss the CAMELS simulation and the data products that we use in this work.  In \S\ref{sec:results_sims}, we present the results of our explorations with the CAMELS simulations, focusing on the information content of the tSZ signal for inferring the impact of feedback on the matter distribution.  In \S\ref{sec:results_data}, we apply our analysis to the DES and ACT measurements.  We summarize our results and conclude in \S\ref{sec:conclusion}.

\section{CAMELS simulations and observables}
\label{sec:camels}

\subsection{Overview of CAMELS simulations}

\begin{table*}
\begin{tabular}{@{}|c|c|l|@{}}
\toprule
Simulation   & Type/Code                                                             & \multicolumn{1}{c|}{\begin{tabular}[c]{@{}c@{}}Astrophysical parameters varied\\ \& its meaning\end{tabular}}                                                                                                                                      \\ \midrule
TNG & \begin{tabular}[c]{@{}c@{}}Magneto-hydrodynamic/\\ AREPO\end{tabular} & \begin{tabular}[c]{@{}l@{}}$A_{\rm SN1}$: (Energy of Galactic winds)/SFR  \\ $A_{\rm SN2}$: Speed of galactic winds\\ $A_{\rm AGN1}$: Energy/(BH accretion rate)\\ $A_{\rm AGN2}$: Jet ejection speed or burstiness\end{tabular}                    \\ \midrule
SIMBA        & Hydrodynamic/GIZMO                                                    & \begin{tabular}[c]{@{}l@{}}$A_{\rm SN1}$ : Mass loading of galactic winds\\ $A_{\rm SN2}$ : Speed of galactic winds\\ $A_{\rm AGN1}$ : Momentum flux in QSO and jet mode of feedback\\ $A_{\rm AGN2}$ : Jet speed in kinetic mode of feedback\end{tabular} \\ \midrule
Astrid       & Hydrodynamic/pSPH                                                     & \begin{tabular}[c]{@{}l@{}}$A_{\rm SN1}$: (Energy of Galactic winds)/SFR  \\ $A_{\rm SN2}$: Speed of galactic winds\\$A_{\rm AGN1}$: Energy/(BH accretion rate)\\ $A_{\rm AGN2}$: Thermal feedback efficiency\end{tabular}                          \\ \bottomrule
\end{tabular}
\caption{Summary of the three feedback models used in this analysis.  For each model, four feedback parameters are varied: $A_{\rm AGN 1}$, $A_{\rm AGN 2}$, $A_{\rm SN 1}$, and $A_{\rm SN 2}$.  The meanings of these parameters are different for each model, and are summarized in the rightmost column.  In addition to these four astrophysical parameters, the cosmological parameters $\Omega_{\rm m}$ and $\sigma_8$ were also varied. \label{tab:feedback}}
\end{table*}

We investigate the use of SZ signals for constraining the impact of feedback on the matter distribution using approximately 3000 cosmological simulations run by the CAMELS collaboration \citep{Villaescusa-Navarro:2021:ApJ:}.  One half of these are gravity-only N-body simulations and the other half are hydrodynamical simulations with matching initial conditions. The simulations are run using three different hydrodynamical sub-grid codes, TNG \citep{Pillepich:2018:MNRAS:}, SIMBA \citep{Dave:2019:MNRAS:} and Astrid \citep{Bird:2022:MNRAS:, Ni:2022:MNRAS:}. As detailed in \citet{Villaescusa-Navarro:2021:ApJ:}, for each sub-grid implementation six parameters are varied: two cosmological parameters ($\Omega_{\rm m}$ and $\sigma_8$) and four parameters dealing with baryonic astrophysics. Of these, two deal with supernovae feedback ($A_{\rm SN1}$ and $A_{\rm SN2}$) and two deal with AGN feedback ($A_{\rm AGN1}$ and $A_{\rm AGN2}$).  The meanings of the feedback parameters for each subgrid model are summarized in Table~\ref{tab:feedback}. 

Note that the astrophysical parameters have somewhat different physical meanings for the different subgrid prescriptions, and there is usually a complex interplay between the parameters and their impact on the properties of galaxies and gas. For example, the parameter $A_{\rm SN1}$ approximately corresponds to the pre-factor for the overall energy output in galactic wind feedback per-unit star-formation in both the TNG \citep{Pillepich:2018:MNRAS:} and Astrid \citep{Bird:2022:MNRAS:} simulations. However, in the SIMBA simulations it corresponds to the to the wind-driven mass outflow rate per unit star-formation calibrated from the Feedback In Realistic Environments (FIRE) zoom-in simulations \citep{Angles-Alcazar:2017:MNRAS:b}. Similarly, the $A_{\rm AGN2}$ parameter controls the burstiness and the temperature of the heated gas during the AGN bursts in the TNG simulations \citep{Weinberger:2017:MNRAS:}.  In the SIMBA suite, it corresponds to the speed of the kinetic AGN jets with constant momentum flux \citep{Angles-Alcazar:2017:MNRAS:a, Dave:2019:MNRAS:}. However, in the Astrid suite, it corresponds to the efficiency of thermal mode of AGN feedback.  As we describe in \S~\ref{sec:fbY}, this can result in counter-intuitive impact on the matter power spectrum in the Astrid simulation, relative to TNG and SIMBA.

For each of the sub-grid physics prescriptions, three varieties of  simulations are provided. These include 27 sims for which the parameters are fixed and initial conditions are varied (cosmic variance, or CV, set), 66 simulations varying only one parameter at a time (1P set)  and 1000 sims varying parameters in a six dimensional latin hyper-cube (LH set). We use the CV simulations to estimate the variance expected in the matter power suppression due to stochasticity (see Fig.~\ref{fig:Pk_Bk_CV}). We use the 1P sims to understand how the  matter suppression responds to variation in each parameter individually. Finally we use the full LH set to effectively marginalize over the full parameter space varying all six parameters. We use publicly available power spectrum and bispectrum measurements for these simulation boxes \citep{Villaescusa-Navarro:2021:ApJ:}.\footnote{See also \url{https://www.camel-simulations.org/data}.} Where unavailable, we calculate the power spectrum and bispectrum, using the publicly available code \texttt{Pylians}.\footnote{\url{https://github.com/franciscovillaescusa/Pylians3}}

\subsection{Baryonic effects on the power spectrum and bispectrum}

\begin{figure*}
\includegraphics[width=\textwidth]{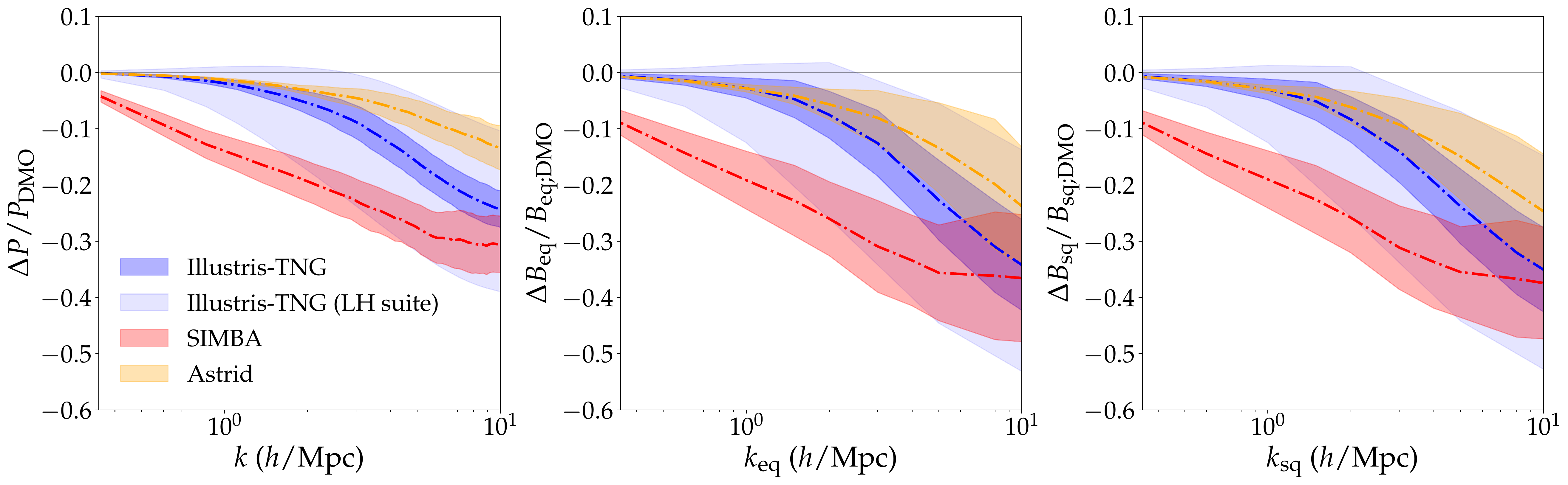}
\caption[]{Far left: Baryonic suppression of the matter power spectrum, $\Delta P/P_{\rm DMO}$, in the CAMELS simulations. 
The dark-blue, red and orange shaded regions correspond to the $1\sigma$ range of the cosmic variance (CV) suite of TNG, SIMBA and Astrid simulations, respectively. The light-blue region corresponds to the $1\sigma$ range associated with the latin hypercube (LH) suite of TNG, illustrating the range of feedback models explored across all parameter values.  Middle and right panels: the impact of baryonic feedback on the matter bispectrum for equilateral and squeezed triangle configurations, respectively.  
}
\label{fig:Pk_Bk_CV}
\end{figure*}

The left panel of Fig.~\ref{fig:Pk_Bk_CV} shows the measurement of the power spectrum suppression caused by baryonic effects in the TNG, SIMBA, and Astrid simulations for their fiducial feedback settings.  The right two panels of the figure show the impact of baryonic effects on the bispectrum for two different tringle configurations (equilateral and squeezed).  To compute these quantitites, we use the matter power spectra and bispectra of the hydrodynamical simulations (hydro) and the dark-matter only (DMO) simulations generated at varying initial conditions (ICs).  For each of the 27 unique IC runs, we calculate the ratios $\Delta P/P_{\rm DMO} = (P_{\rm hydro} - P_{\rm DMO})/P_{\rm DMO}$ and $\Delta B/B_{\rm DMO} = (B_{\rm hydro} - B_{\rm DMO})/B_{\rm DMO}$. As the hydro-dynamical and the N-body simulations are run with same initial conditions, the ratios $\Delta P/P_{\rm DMO}$ and $\Delta B/B_{\rm DMO}$ are roughly independent of sample variance.

It is clear that the amplitude of suppression of the small-scale matter power spectrum can be significant: suppression on the order of tens of percent is reached for all three simulations.  It is also clear that the impact is significantly different between the three simulations.  Even for the simulations in closest agreement (TNG and Astrid), the measurements of $\Delta P/P_{\rm DMO}$ disagree by more than a factor of two at $k = 5\,h/{\rm Mpc}$.  The width of the curves in Fig.~\ref{fig:Pk_Bk_CV} represents the standard deviation measured across the cosmic variance simulations, which all have the same parameter values but different initial conditions. For the bispectrum, we show both the equilateral and squeezed triangle configurations with cosine of angle between long sides fixed to $\mu = 0.9$. Interestingly, the spread in $\Delta P/P_{\rm DMO}$ and $\Delta B/B_{\rm DMO}$ increases with increasing $k$ over the range $0.1 \,h/{\rm Mpc} \lesssim k  \lesssim 10\,h/{\rm Mpc}$.   This increase is driven by stochasticity arising from baryonic feedback.  The middle and right panels show the impact of feedback on the bispectrum for the equilateral and squeezed triangle configurations, respectively.

Throughout this work, we will focus on the regime $0.3\,h/{\rm Mpc}< k < 10\,h/{\rm Mpc}$. Larger scales modes are not present in the $(25 {\rm Mpc}/h)^3$ CAMELS simulations, and in any case, the impact of feedback on large scales is typically small.  Much smaller scales, on the other hand, are difficult to model even in the absence of baryonic feedback \citep{Schneider:2016:JCAP:}. In Appendix~\ref{app:volume_res_comp} we show how the matter power suppression changes when varying the resolution and volume of the simulation boxes. When comparing with the original TNG boxes, we find that while the box sizes do not change the measured power suppression significantly, the resolution of the boxes has a non-negligible impact. This is expected since the physical effect of feedback mechanisms depend on the resolution of the simulations. Note that the errorbars presented in Fig.~\ref{fig:Pk_Bk_CV} will also depend on  the default choice of feedback values assumed.

\subsection{Measuring gas profiles around halos}

We use 3D grids of various fields (e.g. gas density and pressure) made available by the CAMELS team to extract the profiles of these fields around dark matter halos. The grids are generated with resolution of 0.05 Mpc/$h$. 
Following \citet{vanDaalen:2020}, we define $f_b$ as $(M_{\rm gas} + M_{\rm stars})/M_{\rm total}$, where $M_{\rm gas}$, $M_{\rm stars}$ and $M_{\rm  total}$ are the mass in gas, stars and all the components, respectively. 
 The gas mass is computed by integrating the gas number density profile around each halo. 
 We typically measure $f_b$ within the spherical overdensity radius $r_{\rm 500c}$.\footnote{We define spherical overdensity radius ($r_{\Delta c}$, where $\Delta = 200, 500$) and overdensity mass ($M_{\Delta c}$) such that the mean density within $r_{\Delta}$ is $\Delta$ times the critical density $\rho_{\rm crit}$, $M_{\Delta} = \Delta \frac{4}{3} \pi r^3_{\Delta} \rho_{\rm crit}$.} 

The SZ effect is sensitive to the electron pressure.  We compute the electron pressure profiles, $P_e$, using $P_e = 2(X_{\rm H} + 1)/(5X_{\rm H} + 3)P_{\rm th}$, where $P_{\rm th}$ is the total thermal pressure, and $X_{\rm H}= 0.76$ is the primordial hydrogen fraction. Given the electron pressure profile, we measure the integrated SZ signal within $r_{\rm 500c}$ as:
\begin{equation}\label{eq:Y500_from_Pe}
	Y_{\rm 500c} = \frac{\sigma_{\rm T}}{m_e c^2}\int_0^{r_{\rm 500c}} 4\pi r^2 \, P_e(r) \, dr,
\end{equation}
where, $\sigma_{\rm T}$ is the Thomson scattering cross-section, $m_{e}$ is the electron mass and $c$ is the speed of light.

We normalize the SZ observables by the self-similar expectation \citep{Battaglia:2012:ApJ:a},\footnote{Note that we use spherical overdensity mass corresponding to $\Delta = 500$ and hence adjust the coefficients accordingly, while keeping other approximations used in their derivations as the same.} 
\begin{equation}
\label{eq:y_ss}
	Y^{\rm SS} = 131.7 h^{-1}_{70} \,\bigg( \frac{M_{500c}}{10^{15} h^{-1}_{70} M_{\odot}} \bigg)^{5/3} \frac{\Omega_{\rm b}}{0.043} \frac{0.25}{\Omega_{\rm m}} \, {\rm kpc^2},
\end{equation}
where, $M_{200c}$ is mass inside $r_{200c}$ and $h_{70} = h/0.7$.  This calculation, which scales as $M^{5/3}$, assumes hydrostatic equilibrium and that the baryon fraction is equal to cosmic baryonic fraction. Hence deviations from this self-similar scaling provide a probe of the effects of baryonic feedback.  Our final SZ observable is defined as $Y_{500c}/Y^{\rm SS}$.  On the other hand, the amplitude of the pressure profile approximately scales as $M^{2/3}$.  Therefore, when considering the pressure profile as the observable, we factor out a $M^{2/3}$ scaling. 
 
\section{Results I: Simulations}
\label{sec:results_sims}

\subsection{Inferring feedback parameters from $f_b$ and $y$} 
\label{sec:fisher}
We first consider how the halo $Y$ signal can be used to constrain the parameters describing the subgrid physics models.  This question has been previously investigated using the CAMELS simulations by \citet{Moser:2022} and \citet{Wadekar:2022}.   The rest of our analysis will focus on constraining changes to the power spectrum and bispectrum, and our intention here is mainly to provide a basis of comparison for those results.  

Similar to \citet{Wadekar:2022}, we treat the mean $\log(Y_{500c}/M^{5/3})$ value of all the halos in two mass bins ($10^{12} < M (M_{\odot}/h) < 5\times 10^{12}$ and $5 \times 10^{12} < M (M_{\odot}/h) < 10^{14}$) as our observable; we refer to this observable as $\vec{d}$.  In this section, we restrict our analysis to only the TNG simulations.  Here and throughout our investigations with CAMELS we ignore the contributions of measurement uncertainty since our intention is mainly to assess the information content of the SZ signals.  We therefore use the CV simulations to determine the covariance, $\mathbf{C}$, of the $\vec{d}$.  Note that the level of cosmic variance will depend on the volume probed, and can be quite large for the CAMELS simulations.  Given this covariance, we use the Fisher matrix formalism to forecast the precision with which the feedback and cosmological parameters can be constrained.  

The Fisher matrix, $F_{ij}$, is given by
\begin{equation}
F_{ij} = \frac{\partial \vec{d}^T}{\partial \theta_i} \mathbf{C}^{-1} \frac{\partial \vec{d}}{\partial \theta_i},
\end{equation}
where $\theta_i$ refers to the $i$th parameter value.  Calculation of the derivatives $\partial \vec{d}/\partial \theta_i$ is complicated by the large amount of stochasticity between the CAMELS simulations.  To perform the derivative calculation, we use a radial basis function interpolation method based on \citet{Moser:2022,Cromer:2022}.  We show an example of the derivative calculation in Appendix~\ref{app:emulation}.  We additionally assume a Gaussian prior on parameter $p$ with $\sigma(\ln p) = 1$ for the feedback parameters and $\sigma(p) = 1$ for the cosmological parameters.  The forecast parameter covariance matrix, $\mathbf{C}_p$, is then related to the Fisher matrix by $\mathbf{C}_p = \mathbf{F}^{-1}$.

\begin{figure}
    \centering
    \includegraphics[scale=0.5]{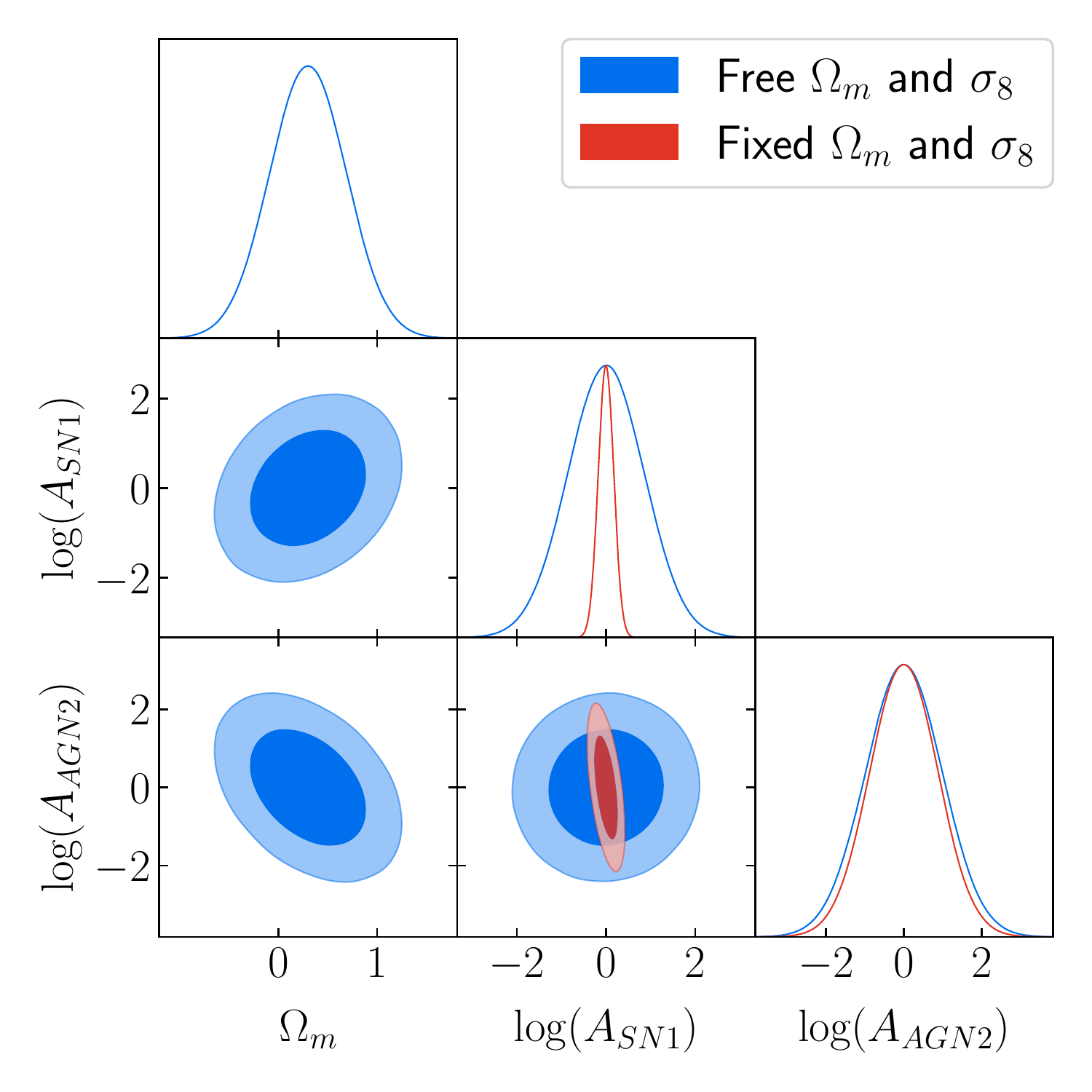}
    \caption{Forecast constraints on the feedback parameters  when $\log Y_{500c}/Y^{\rm SS}$ in two halo mass bins is treated as the observable.  Even when the cosmological model is fixed (red contours), the AGN parameters (e.g. $A_{AGN2}$) remain effectively unconstrained (note that we impose a Gaussian prior with $\sigma(\ln p) = 1$ on all feedback parameters, $p$).  When the cosmological model is free (blue contours), all feedback parameters are unconstrained.  We assume that the only contribution to the variance of the observable is sample variance coming from the finite volume of the CAMELS simulations. 
    }
    \label{fig:fisher}
\end{figure}

The parameter constraints corresponding to our calculated Fisher matrix are shown in Fig.~\ref{fig:fisher}.  We show results only for $\Omega_{\rm m}$, $A_{\rm SN1}$ and $A_{\rm AGN2}$, but additionally marginalize over $\sigma_8$, $A_{\rm SN2}$ and $A_{\rm AGN1}$. The degeneracy directions seen in our results are consistent with those in \citet{Wadekar:2022}.  We we find a weaker constraint on $A_{\rm AGN2}$, likely owing to the large sample variance contribution to our calculation.  

It is clear from Fig.~\ref{fig:fisher} that the marginalized constraints on the feedback parameters are weak. If information about $\Omega_{\rm m}$ is not used, we effectively have no information about the feedback parameters.  Even when $\Omega_{\rm m}$ is fixed, the constraints on the feedback parameters are not very precise.  This finding is consistent with \citet{Wadekar:2022}, for which measurement uncertainty was the main source of variance rather than sample variance. Part of the reason for the poor constraints is the degeneracy between the AGN and SN parameters.  As we show below, the impacts of SN and AGN feedback can have opposite impacts on the $Y$ signal; moreover, even $A_{\rm AGN1}$ and $A_{\rm AGN2}$ can have opposite impacts on $Y$.  These degeneracies, as well as degeneracies with cosmological parameters like $\Omega_m$, make it difficult to extract tight constraints on the feedback parameters from measurements of $Y$.  However, for the purposes of cosmology, we are ultimately most interested in the impact of feedback on the matter distribution, and not the values of the feedback parameters themselves.   These considerations motivate us to  instead explore direct inference of changes to the statistics of the matter distribution from the $Y$ observables.  This will be the focus of the rest of the paper.

\subsection{$f_b$ and $y$ as probes of baryonic effects on the matter power spectrum}
\label{sec:fbY}

As discussed above, \citet{vanDaalen:2020} observed a tight correlation between suppression of the matter power spectrum and the baryon fraction, $f_b$, in halos with $6\times 10^{13} M_{\odot} \lesssim M_{500c} \lesssim 10^{14}\,M_{\odot}$. That relation was found to hold regardless of the details of the feedback implementation, suggesting that by measuring $f_b$, one could robustly infer the impact of baryonic feedback on the power spectrum. We begin by investigating the connection between matter power spectrum suppression and integrated tSZ parameter in low-mass, $M \sim 10^{13}\,M_{\odot}$, halos to test if similar correlation exists (c.f. \citet{Delgado:23} for a similar figure between $f_b$ and $\Delta P/P_{\rm DMO}$). We also consider a wider range of feedback models than \citet{vanDaalen:2020}, including the SIMBA and Astrid models.

\begin{figure}
\includegraphics[width=0.95\columnwidth]{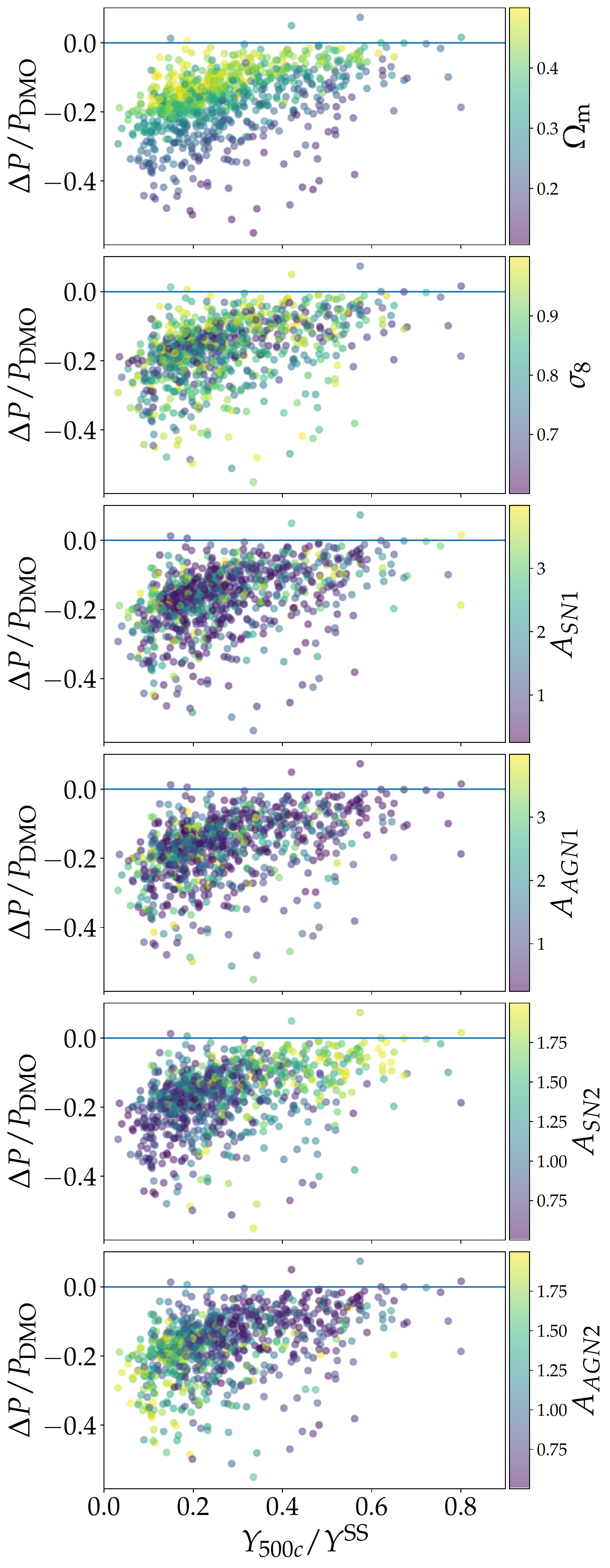}
\caption[]{We show the relation between matter power suppression at $k=2 h/{\rm Mpc}$ and the integrated tSZ signal, $Y_{500c}/Y^{\rm SS}$, of halos in the mass range $10^{13} < M\,(M_{\odot}/h) < 10^{14}$ in the SIMBA simulation suite. In each of six panels, the points are colored corresponding to the parameter value given in the associated colorbar. 
}
\label{fig:Pk_SIMBA_allparams}
\end{figure}

Fig.~\ref{fig:Pk_SIMBA_allparams} shows the impact of cosmological and feedback parameters on the relationship between the power spectrum suppression ($\Delta P/P_{\rm DMO}$) and the ratio $Y_{\rm 500c}/Y^{\rm SS}$
 for the SIMBA simulations.  Each point corresponds to a single simulation, taking the average over all halos with $10^{13} < M (M_{\odot}/h) < 10^{14}$ when computing $Y_{\rm 500c}/Y^{\rm SS}$. Note that since the halo mass function rapidly declines at high masses, the average will be dominated by the low mass halos. We observe that the largest suppression (i.e. more negative $\Delta P/P_{\rm DMO}$) occurs when $A_{\rm AGN2}$ is large.  This is caused by powerful AGN jet-mode feedback ejecting gas from halos, leading to a significant reduction in the matter power spectrum, as described by e.g. \citet{vanDaalen:2020, Borrow:2020:MNRAS:, Gebhardt:23}.  For SIMBA, the parameter $A_{\rm AGN2}$ controls the velocity of the ejected gas, with higher velocities (i.e. higher $A_{\rm AGN2}$) leading to gas ejected to larger distances.  On the other hand, when $A_{\rm SN2}$ is large, $\Delta P/P_{\rm DMO}$ is small.  
 This is because efficient supernovae feedback prevents the formation of massive galaxies which host AGN and hences reduces the strength of the AGN feedback.
 The parameter $A_{\rm AGN1}$, on the other hand, controls the radiative quasar mode of feedback, which has slower gas outflows and thus a smaller impact on the matter distribution.

It is also clear from Fig.~\ref{fig:Pk_SIMBA_allparams} that increasing $\Omega_{\rm m}$ reduces $|\Delta P/P_{\rm DMO}|$, relatively independently of the other parameters. By increasing $\Omega_{\rm m}$, the ratio  $\Omega_{\rm b}/\Omega_{\rm m}$ decreases, meaning that halos of a given mass have fewer baryons, and the impact of feedback is therefore reduced.  We propose a very simple toy model for this effect in  \S\ref{sec:simple_model}.  

The impact of $\sigma_8$ in Fig.~\ref{fig:Pk_SIMBA_allparams} is less clear.  For halos in the mass range shown, we find that increasing $\sigma_8$ leads to a roughly monotonic decrease in $Y_{500c}$ (and $f_b$), presumably because higher $\sigma_8$ means that there are more halos amongst which the same amount of baryons must be distributed.  This effect would not occur for cluster-scale halos because these are rare and large enough to gravitationally dominate their local environments, giving them $f_b \sim \Omega_{\rm b}/\Omega_{\rm m}$, regardless of $\sigma_8$.  In any case, no clear trend with $\sigma_8$ is seen in Fig.~\ref{fig:Pk_SIMBA_allparams} because $\sigma_8$ does not correlate strongly with $\Delta P/P_{\rm DMO}$. 

\begin{figure*}
\includegraphics[width=0.95\textwidth]{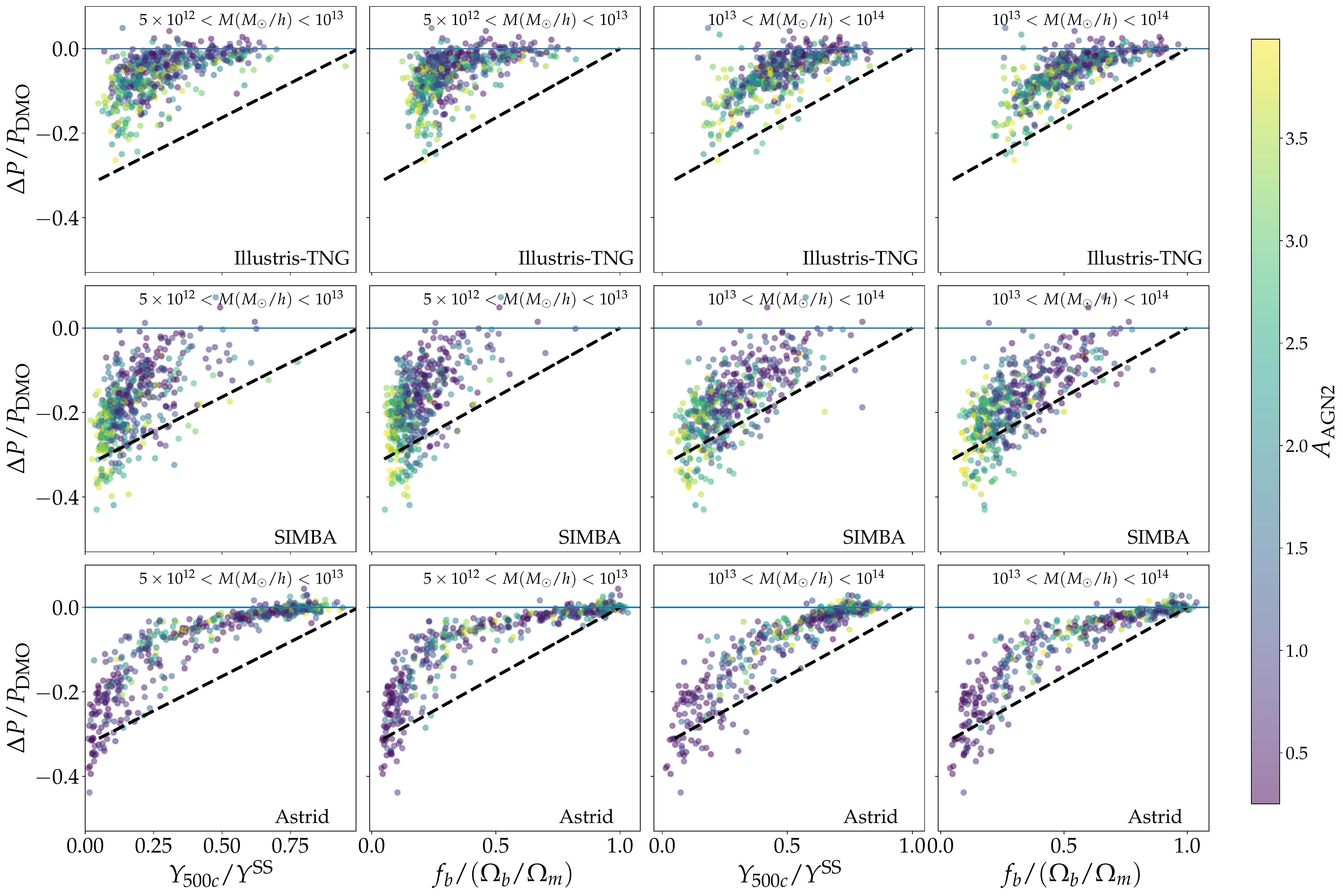}
\caption[]{Impact of baryonic physics on the matter power spectrum at $k=2 h/{\rm Mpc}$ for the TNG, SIMBA and Astrid simulations (top, middle, and bottom rows).  Each point corresponds to an average across halos in the indicated mass ranges in a different CAMELS simulation.  We restrict the figure to simulations that have $0.2 < \Omega_{\rm m} < 0.4$.   The dashed curves illustrate the behavior of the model described in \S\ref{sec:simple_model} when the gas ejection distance is large compared to the halo radius and $2\pi/k$. 
}
\label{fig:Y_fb_DeltaP}
\end{figure*}

Fig.~\ref{fig:Y_fb_DeltaP} shows the relationship between $\Delta P/P_{\rm DMO}$ at  $k = 2\,h/{\rm Mpc}$ and $f_b$ or $Y_{500c}$ in different halo mass bins and for different amounts of feedback, colored by the value of $A_{\rm AGN2}$.  As in Fig.~\ref{fig:Pk_SIMBA_allparams}, each point represents an average over all halos in the indicated mass range for a particular CAMELS simulation (i.e. at fixed values of cosmological and feedback parameters).    Note that the meaning of $A_{\rm AGN2}$ is not exactly the same across the different feedback models, as noted in \S\ref{sec:camels}. For TNG and SIMBA we expect increasing $A_{\rm AGN2}$ to lead to stronger AGN feedback driving more gas out of the halos, leading to more power suppression without strongly regulating the growth of black holes. However, for Astrid, increasing $A_{\rm AGN2}$ parameter would more strongly regulate and suppress the black hole growth in the box since controls the efficiency of thermal mode of AGN feedback \citep{Ni:2022:MNRAS:}. This drastically reduces the number of high mass black holes and hence effectively reducing the amount of feedback that can push the gas out of the halos, leading to less matter power suppression. We see this difference reflected in Fig.~\ref{fig:Y_fb_DeltaP} where for the Astrid simulations the points corresponding to high $A_{\rm AGN2}$, result in $\Delta P/P_{\rm DMO} \sim 0$, in contrast to TNG and SIMBA suite of simulations.

For the highest mass bin ($10^{13} < M (M_{\odot}/h) < 10^{14}$, rightmost column of Fig.~\ref{fig:Y_fb_DeltaP}) our results are in agreement with \citet{vanDaalen:2020}: we find that there is a robust correlation between between $f_b/(\Omega_{\rm b}/\Omega_{\rm m})$ and the matter power suppression (also see \citet{Delgado:23}). This relation is roughly consistent across different feedback subgrid models, although the different  models appear to populate different parts of this relation.  Moreover, varying $A_{\rm AGN2}$ appears to move points along this relation, rather than broadening the relation.  This is in contrast to $\Omega_{\rm m}$, which as shown in Fig.~\ref{fig:Pk_SIMBA_allparams}, tends to move simulations in the direction orthogonal to the narrow $Y_{500c}$-$\Delta P/P_{\rm DMO}$ locus.  For this reason, and given current constraints on $\Omega_{\rm m}$, we restrict Fig.~\ref{fig:Y_fb_DeltaP} to simulations with $0.2 < \Omega_{\rm m} < 0.4$. 
 The dashed curves shown in Fig.~\ref{fig:Y_fb_DeltaP} correspond to the toy model discussed in \S\ref{sec:simple_model}.

At low halo mass, the relation between $f_b/(\Omega_{\rm b}/\Omega_{\rm m})$ and $\Delta P/P_{\rm DMO}$ appears similar to that for the high-mass bin, although it is somewhat flatter at high $f_b$, and somewhat steeper at low $f_b$.  Again the results are fairly consistent across the different feedback prescriptions, although points with high $f_b/(\Omega_{\rm b}/\Omega_{\rm m})$ are largely absent for SIMBA. This is because the feedback mechanisms are highly efficient in SIMBA, driving the gas out of their parent halos.  

The relationships between $Y$ and $\Delta P/P_{\rm DMO}$ appear quite similar to those between $\Delta P/P_{\rm DMO}$ and $f_b/(\Omega_{\rm b}/\Omega_{\rm m})$. This is not too surprising because $Y$ is sensitive to the gas density, which dominates $f_b/(\Omega_{\rm b}/\Omega_{\rm m})$.  However, $Y$ is also sensitive to the gas temperature.  Our results suggest that variations in gas temperature are not significantly impacting the $Y_{500c}$-$\Delta P/P_{\rm DMO}$ relation.  The possibility of using the tSZ signal to infer the impact of feedback on the matter distribution rather than $f_b/(\Omega_{\rm b}/\Omega_{\rm m})$ is therefore appealing.  This will be the focus of the remainder of the paper.

\begin{figure*}
\includegraphics[width=0.95\textwidth]{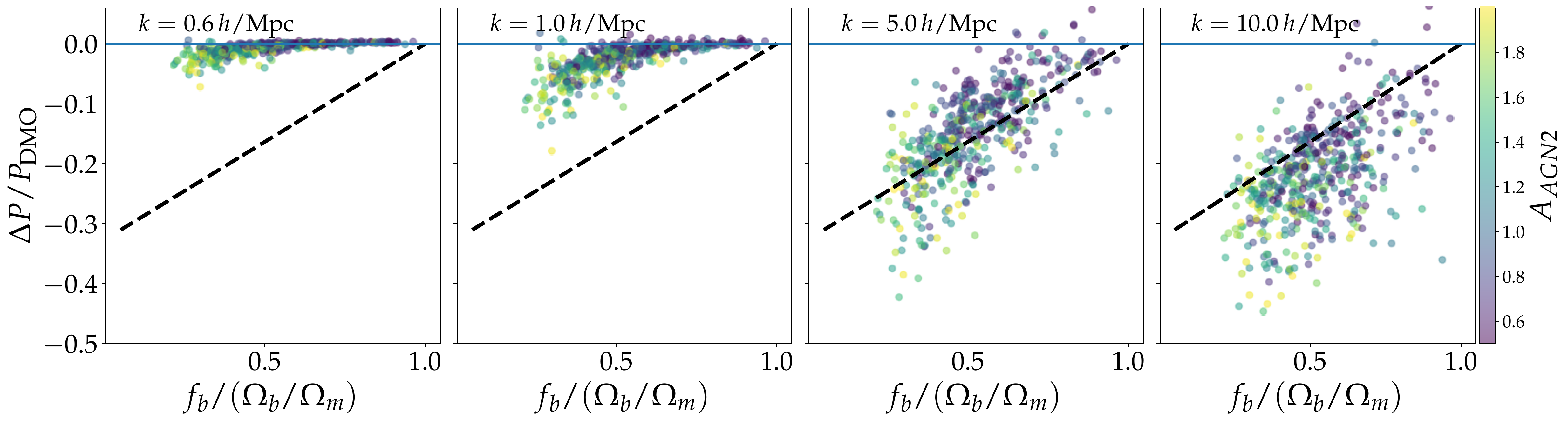}
\caption[]{Similar to Fig.~\ref{fig:Y_fb_DeltaP}, but for different values of $k$.  For simplicity, we show only the TNG simulations for halos in the mass range $10^{13} < M (M_{\odot}/h) < 10^{14}$.  The dashed curves illustrate the behavior of the model described in \S\ref{sec:simple_model} in the regime that the radius to which gas is ejected by AGN is larger than the halo radius, and larger than $2\pi/k$.  As expected, this model performs best in the limit of high $k$ and large halo mass. 
}
\label{fig:scatter_plot_all_ks}
\end{figure*}

Fig.~\ref{fig:scatter_plot_all_ks} shows the same quantities as Fig.~\ref{fig:Y_fb_DeltaP}, but now for a fixed halo mass range ($10^{13} < M/(M_{\odot}/h) < 10^{14}$), fixed subgrid prescription (TNG), and varying values of $k$.  We find roughly similar results when using the different subgrid physics prescriptions.  At low $k$, we find that there is a regime at high $f_b/(\Omega_{\rm b}/\Omega_{\rm m})$ for which $\Delta P /P_{\rm DMO}$ changes negligibly.  It is only when $f_b/(\Omega_{\rm b}/\Omega_{\rm m})$ becomes very low that $\Delta P/P_{\rm DMO}$ begins to change.  On the other hand, at high $k$, there is a near-linear relation between $f_b/(\Omega_{\rm b}/\Omega_{\rm m})$ and $\Delta P/P_{\rm DMO}$.  

\subsection{A toy model for power suppression}
\label{sec:simple_model}

We now describe a simple model for the effects of feedback on the relation between $f_b$ or $Y$ and $\Delta P/P_{\rm DMO}$ that explains some of the features seen in Figs.~\ref{fig:Pk_SIMBA_allparams}, \ref{fig:Y_fb_DeltaP} and \ref{fig:scatter_plot_all_ks}.  We assume in this model that it is removal of gas from halos by AGN feedback that is responsible for changes to the matter power spectrum.   SN feedback, on the other hand, can prevent gas from accreting onto the SMBH and therefore reduce the impact of AGN feedback \citep{Angles-Alcazar:2017:MNRAS:c, Habouzit:2017:MNRAS:}.  This scenario is consistent with the fact that at high SN feedback, we see that $\Delta P/P_{\rm DMO}$ goes to zero (second panel from the bottom in Fig.~\ref{fig:Pk_SIMBA_allparams}).  Stellar feedback can also prevent gas from accreting onto low-mass halos \citep{Pandya:2020:ApJ:, Pandya:2021:MNRAS:}.  In some sense, the distinction between gas that is ejected by AGN and gas that is prevented from accreting onto halos by stellar feedback does not matter for our simple model.  Rather, all that matters is that some amount of gas that would otherwise be in the halo is instead outside of the halo as a result of feedback effects, and  it is this gas which is responsible for changes to the matter power spectrum. 

We identify three relevant scales: (1) the halo radius, $R_h$,  (2) the distance to which gas is ejected by the AGN, $R_{\rm ej}$, and (3) the scale at which the power spectrum is measured, $2\pi/k$.    If $R_{\rm ej} \ll 2\pi/k$, then there will be no impact on $\Delta P$ at $k$: this corresponds to a rearrangement of the matter distribution on scales well below where we measure the power spectrum.  If, on the other hand, $R_{\rm ej} \ll R_h$, then there will be no impact on $f_b$ or $Y$, since the gas is not ejected out of the halo.  We therefore consider four regimes defined by the relative amplitudes of $R_h$, $R_{\rm ej}$, and $2\pi/k$, as described below. Note that there is not a one-to-one correspondence between physical scale in configuration space and $2\pi/k$; therefore, the inequalities below should be considered as approximate.  The four regimes are:
\begin{itemize}
\item Regime 1: $R_{\rm ej} < R_h$ and $R_{\rm ej} < 2\pi /k$. In this regime, changes to the feedback parameters have no impact on $f_b$ or $\Delta P$. 
\item Regime 2: $R_{\rm ej} > R_h$ and $R_{\rm ej} < 2\pi/k$.  In this regime, changes to the feedback parameters result in movement along the $f_b$ or $Y$ axis without changing $\Delta P$. Gas is being removed from the halo, but the resultant changes to the matter distribution are below the scale at which we measure the power spectrum.  Note that  Regime 2 cannot occur when $R_h > 2\pi/k$ (i.e.  high-mass halos at large $k$). 
\item Regime 3: $R_{\rm ej} > R_h$ and $R_{\rm ej} > 2\pi/k$.  In this regime, changing the feedback amplitude directly changes the amount of gas ejected from halos as well as $\Delta P/P_{\rm DMO}$.  
\item Regime 4: $R_{\rm ej} < R_h$ and $R_{\rm ej} > 2 \pi/k$.  In this regime, gas is not ejected out of the halo, so $f_b$ and $Y$ should not change.  In principle, the redistribution of gas within the halo could lead to changes in $\Delta P/P_{\rm DMO}$.  However, as we discuss below, this does not seem to happen in practice.
\end{itemize}

Let us now consider the behavior of $\Delta P/P_{\rm DMO}$ and $f_b$ or $Y$ as the feedback parameters are varied in Regime 3.   A halo of mass $M$ is associated with an overdensity $\delta_m$ in the absence of feedback, which is changed to $\delta'_m$ due to ejection of baryons as a result of feedback.  In Regime 3, some amount of gas, $M_{\rm ej}$, is completely removed from the halo.  This changes the size of the overdensity associated with the halo to \begin{eqnarray}
    \frac{\delta_m'}{\delta_m} &=& 1 - \frac{M_{\rm ej}} {M}.
\end{eqnarray}
The change to the power spectrum is then
\begin{eqnarray}
\label{eq:deltap_over_p}
    \frac{\Delta P}{P_{\rm DMO}} &\sim& \left(  \frac{\delta_m'}{\delta_m} \right)^2  -1 \approx -2\frac{M_{\rm ej}}{M},
\end{eqnarray}
where we have assumed that $M_{\rm ej}$ is small compared to $M$.  We have ignored the $k$ dependence here, but in Regime 3, the ejection radius is larger than the scale of interest, so the calculated $\Delta P/P_{\rm DMO}$ should apply across a range of $k$ in this regime.

The ejected gas mass can be related to the gas mass in the absence of feedback.  We  write the gas mass in the absence of feedback as  $f_c (\Omega_{\rm b}/\Omega_{\rm m}) M$, where $f_c$ encapsulates non-feedback processes that result in the halo having less than the cosmic baryon fraction.  We then have
\begin{eqnarray}
    M_{\rm ej} &=& f_c(\Omega_{\rm b}/\Omega_{\rm m})M  - f_{b} M - M_0,
\end{eqnarray}
where $M_0$ is the mass that has been removed from the gaseous halo, but that does not change the power spectrum, e.g. the conversion of gas to stars.  Substituting into Eq.~\ref{eq:deltap_over_p}, we have
\begin{eqnarray}\label{eq:DelP_P_fb}
    \frac{\Delta P}{P_{\rm DMO}} = -2 \frac{f_c\Omega_{\rm b}}{\Omega_{\rm m}} \left( 1 -\frac{f_{b}\Omega_{\rm m}}{f_c \Omega_{\rm b}}  - \frac{\Omega_{\rm m} M_0}{f_c \Omega_{\rm b} M} \right).
\end{eqnarray}
In other words, for Regime 3, we find a linear relation between $\Delta P/P_{\rm DMO}$ and $f_b \Omega_{\rm m}/\Omega_{\rm b}$.  For high mass halos, we should have $f_c \approx 1$ and $M_0/M \approx 0$.  In this limit, the relationship between $f_b$ and $\Delta P/P_{\rm DMO}$ becomes
\begin{eqnarray}\label{eq:DelP_P_fb_2}
    \frac{\Delta P}{P_{\rm DMO}} = -2 \frac{\Omega_{\rm b}}{\Omega_{\rm m}} \left( 1 -\frac{f_{b}\Omega_{\rm m}}{\Omega_{\rm b}}  \right),
\end{eqnarray}
which is linear between endpoints at $(\Delta P/P_{\rm DMO},f_b \Omega_{\rm m}/\Omega_{\rm b}) = (-2\Omega_{\rm b}/\Omega_{\rm m},0)$ and $(\Delta P/P_{\rm DMO},f_b \Omega_{\rm m}/\Omega_{\rm b}) = (0,1)$.  We show this relation as the dashed line in the $f_b$ columns of Figs.~\ref{fig:Y_fb_DeltaP} and Fig.~\ref{fig:scatter_plot_all_ks}.

We can repeat the above argument for $Y$.  Unlike the case with $f_b$, processes other than the removal of gas may reduce $Y$; these include, e.g., changes to the gas temperature in the absence of AGN feedback, or nonthermal pressure support.  We account for these with a term $Y_0$, defined such that when $M_{\rm ej} = M_0 = 0$, we have $Y + Y_0 = f_c (\Omega_{\rm b}/\Omega_{\rm m}) MT /\alpha$, where we have assumed constant gas temperature, $T$, and $\alpha$ is a dimensionful constant of proportionality.   We ignore detailed modeling of variation in the temperature of the gas due to feedback and departures from hydro-static equilibrium \citep{Ostriker:2005:ApJ:}.  We then have
\begin{eqnarray} 
\frac{\alpha(Y+Y_0)}{T} = f_c (\Omega_{\rm b} / \Omega_{\rm m})M - M_{\rm ej} - M_0.
\end{eqnarray}
Substituting the above equation into Eq.~\ref{eq:deltap_over_p} we have
\begin{eqnarray}
    \frac{\Delta P}{P_{\rm DMO}} &=& 
 -2\frac{f_c\Omega_{\rm b}}{\Omega_{\rm m}} \left(1 - \frac{\alpha (Y+Y_0) \Omega_{\rm m}}{f_c TM \Omega_{\rm b}} - \frac{\Omega_{\rm m} M_0}{f_c \Omega_{\rm b} M} \right) . \nonumber \\
\end{eqnarray}
Following Eq.~\ref{eq:y_ss}, we define the self-similar value of $Y$, $Y^{\rm SS}$, via
\begin{eqnarray}
\alpha Y^{\rm SS}/T = (\Omega_{\rm b}/\Omega_{\rm m})M, 
\end{eqnarray}
leading to
\begin{eqnarray}
    \frac{\Delta P}{P_{\rm DMO}} &=& -2\frac{f_c\Omega_{\rm b}}{\Omega_{\rm m}} \left(1 - \frac{(Y+Y_0)}{f_c Y^{\rm SS}}  - \frac{\Omega_{\rm m} M_0}{f_c \Omega_{\rm b} M}\right).
\end{eqnarray}
Again taking the limit that $f_c \approx 1$ and $M_0/M \approx 0$, we have
\begin{eqnarray}
    \frac{\Delta P}{P_{\rm DMO}} &=& -2\frac{\Omega_{\rm b}}{\Omega_{\rm m}} \left(1 - \frac{(Y+Y_0)}{ Y^{\rm SS}}  \right).
\end{eqnarray}
Thus, we see that in Regime 3, the relation between $Y/Y^{\rm SS}$ and $\Delta P/P_{\rm DMO}$ is linear.  The $Y/Y^{\rm SS}$ columns of Figs.~\ref{fig:Y_fb_DeltaP} show this relationship, assuming $Y_0 = 0$.

In summary, we interpret the results of Figs.~\ref{fig:Y_fb_DeltaP} and \ref{fig:scatter_plot_all_ks} in the following way.  Starting at low feedback amplitude, we are initially in Regime 1.  In this regime, the simulations cluster around $f_b f_c \Omega_{\rm m}/\Omega_{\rm b} \approx 1$ (or $Y \approx Y_0$) and $\Delta P/P_{\rm DMO} \approx 0$ since changing the feedback parameters in this regime does not impact $f_b$ or $\Delta P/P_{\rm DMO}$.  For high mass halos, we have $f_c \approx 1$ and $Y_0 \approx 0$ (although SIMBA appears to have $Y_0 >0$, even at high mass); for low mass halos, $f_c < 1$ and $Y_0 >0$.  As we increase the AGN feedback amplitude, the behavior is different depending on halo mass and $k$:
\begin{itemize}
\item For low halo masses or low $k$, increasing the AGN feedback amplitude  leads the simulations into Regime 2.  Increasing the feedback amplitude in this regime moves points to lower $Y/Y^{\rm SS}$ (or $f_b \Omega_{\rm m}/\Omega_{\rm b}$) without significantly impacting $\Delta P/P_{\rm DMO}$.   Eventually, when the feedback amplitude is sufficiently strong, these halos enter Regime 3, and we see a roughly linear decline in $\Delta P/P_{\rm DMO}$ with decreasing $Y/Y^{\rm SS}$ (or $f_b\Omega_{\rm m}/\Omega_{\rm b}$), as discussed above.  
\item For high mass halos and high $k$, we never enter Regime 2 since it is not possible to have $R_{\rm ej} > R_h$ and $R_{\rm ej} < 2\pi/k$ when $R_h$ is very large.  In this case, we eventually enter Regime 3, leading to a linear trend of decreasing $\Delta P/P_{\rm DMO}$ with decreasing $Y/Y^{\rm SS}$ or $f_b \Omega_{\rm m}/\Omega_{\rm b}$, as predicted by the above discussion.  This behavior is especially clear in Fig.~\ref{fig:scatter_plot_all_ks}: at high $k$, the trend closely follows the predicted linear relation.  At low $k$, on the other hand, we see a more prominent Regime 2 region.  The transition between these two regimes is expected to occur when $k \sim 2\pi/R_h$, which is roughly $5\,h^{-1}{\rm Mpc}$ for the halo mass regime shown in the figure.  This expectation is roughly confirmed in the figure. 
\end{itemize}
Interestingly, we never see Regime 4 behavior: when the halo mass is large and $k$ is large, we do not see rapid changes in $\Delta P/P_{\rm DMO}$ with little change to $f_b$ and $Y$.  This could be because this regime corresponds to movement of the gas entirely within the halo.  If the gas has time to re-equilibrate, it makes sense that we would see little change to $\Delta P/P_{\rm DMO}$ in this regime.

\subsection{Predicting the power spectrum suppression from the halo observables}

While the toy model described above roughly captures the trends between $Y$ (or $f_b$) and $\Delta P/P_{\rm DMO}$, it of course does not capture all of the physics associated with feedback.  It is also clear that there is significant scatter in the relationships between observable quantities and $\Delta P$.  It is possible that this scatter is reduced in some higher dimensional space that includes more observables.  To address both of these issues, we now train statistical models to learn the relationships between observable quantities and $\Delta P/P_{\rm DMO}$.  We will focus on results obtained with random forest regression \citep{Breiman2001}.  We have also tried using neural networks to infer these relationships, but have not found any significant improvement with respect to the random forest results, presumably because the space is low-dimensional (i.e. we consider at most about five observable quantities at a time). We leave a detailed comparison with other decision tree based approaches, such as gradient boosted trees \citep{Friedman_boosted_tree:01} to a future study.

\begin{figure*}
\includegraphics[width=0.95\textwidth]{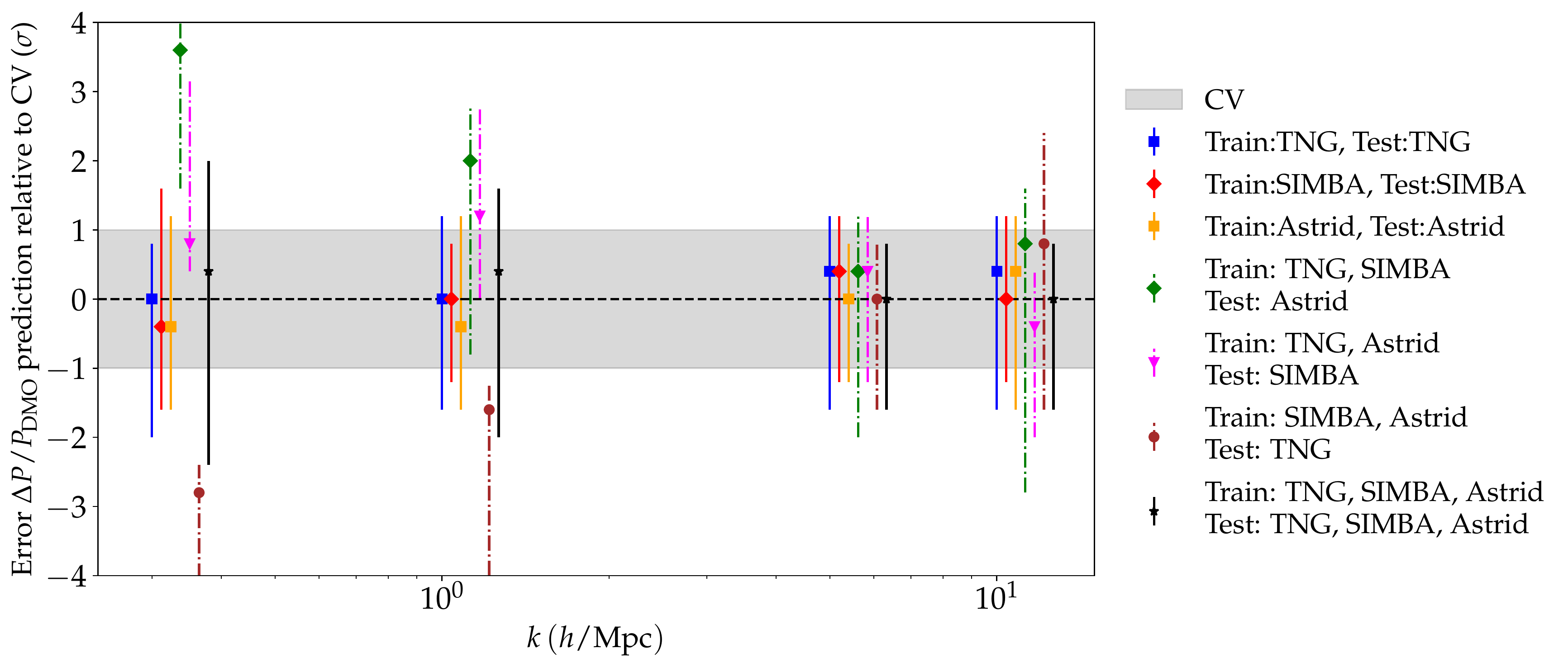}
\caption[]{
We show the results of the random forest regressor predictions for the baryonic power suppression, represented by $\Delta P/P_{\rm DMO}$, across the LH suite of simulations at four different scales $k$ using the subgrid physics models for TNG, SIMBA, and Astrid. The model was trained using the average $f_b$ of halos with masses between $5\times10^{12} < M (M_{\odot}/h) < 10^{14}$ and the cosmological parameter $\Omega_{\rm m}$. The errorbars indicate the uncertainty in the predictions normalized by the uncertainity in the CV suite at each scale, showing the 16-84 percentile error on the test set. The gray band represents the expected 1$\sigma$ error from the CV suite.  The model performs well when the training and test simulations are the same. When tested on an independent simulation, it remains robust at high $k$ but becomes biased at low $k$. The results presented in the remainder of the paper are based on training the model on all three simulations. The data points at each scale are staggered for clarity.
}
\label{fig:Pk_Y_CV}
\end{figure*}

We train a random forest model to go from observable quantities (e.g. $f_b/(\Omega_{\rm b}/\Omega_{\rm m})$ and $Y_{500c}/Y^{\rm SS}$) to a prediction for $\Delta P/P_{\rm DMO}$ at multiple $k$ values. The random forest model uses 100 trees with a ${\rm max}_{\rm depth} = 10$.\footnote{We use a publicly available code: \url{https://scikit-learn.org/stable/modules/generated/sklearn.ensemble.RandomForestRegressor.html}. We also verified that our conclusions are robust to changing the settings of the random forest.} In this section we analyze the halos in the mass bin $5\times 10^{12} < M_{\rm halo} (M_{\odot}/h) < 10^{14}$ but we also show the results for halos with lower masses in Appendix~\ref{app:low_mass}.  We also consider supplying the values of $\Omega_{\rm m}$ as input to the random forest, since it can be constrained precisely through other observations (e.g. primary CMB 
observations), and as we showed in \S\ref{sec:fbY}, the cosmological parameters can impact the observables.\footnote{One might worry that using cosmological information to constrain $\Delta P/P_{\rm DMO}$ defeats the whole purpose of constraining $\Delta P/P_{\rm DMO}$ in order to improve cosmological constraints.  However, observations, such as those of CMB primary anisotropies, already provide precise constraints on the matter density without using information in the small-scale matter distribution.  
}

Ultimately, we are interested in making predictions for $\Delta P/P_{\rm DMO}$ using observable quantities.  However, the sample variance in the CAMELS simulations limits the precision with which we can measure $\Delta P/P_{\rm DMO}$.  It is not possible to predict $\Delta P/P_{\rm DMO}$ to better than this precision. We will therefore normalize the uncertainties in the RF predictions by the cosmic variance error. In order to obtain the uncertainty in the predictions, we randomly split the data into 70\% training and 30\% test set. After training the RF regressor using the training set and a given observable, we make compute the 16th and 84th percentile of the distribution of prediction errors evaluated on the test set.  This constitutes our assessment of prediction uncertainty.  

\begin{figure*}
    \centering
    \includegraphics[width=0.95\textwidth]{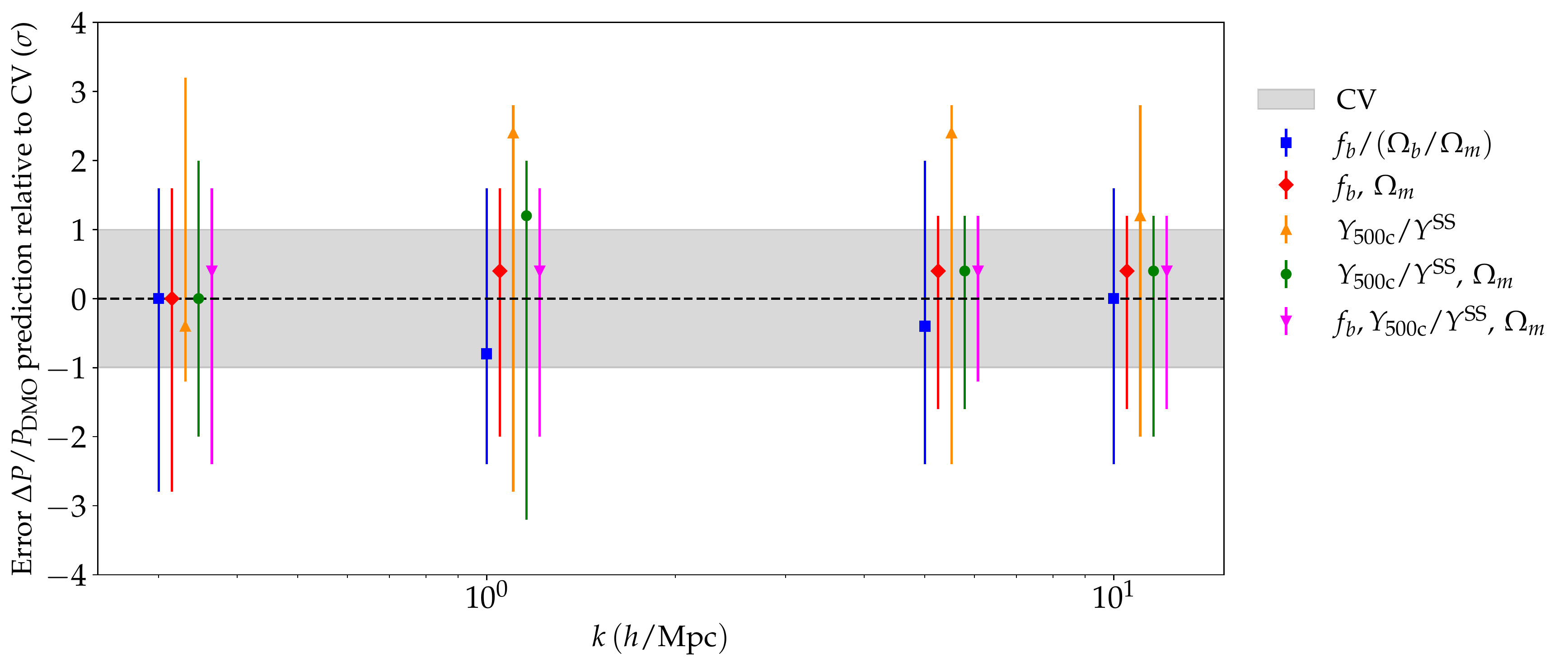}    
    \caption{
    Similar to Fig.~\ref{fig:Pk_Y_CV}, but showing results 
    when training the RF model on different observables from all three simulations (TNG, SIMBA and Astrid) to predict $\Delta P/P_{\rm DMO}$ of a random subset of the the three simulations not used in training. 
    We find that jointly training on the deviation of the integrated SZ profile from the self-similar expectation, $Y_{500c}/Y^{\rm SS}$ and $\Omega_{\rm m}$ results in inference of power suppression that is comparable to cosmic variance errors, with small improvements when additionally adding the baryon fraction ($f_b$) of halos in the above mass range.  
    }
    \label{fig:predict_y500_fb}
\end{figure*}

\begin{figure*}
    \centering
    \includegraphics[width=0.95\textwidth]{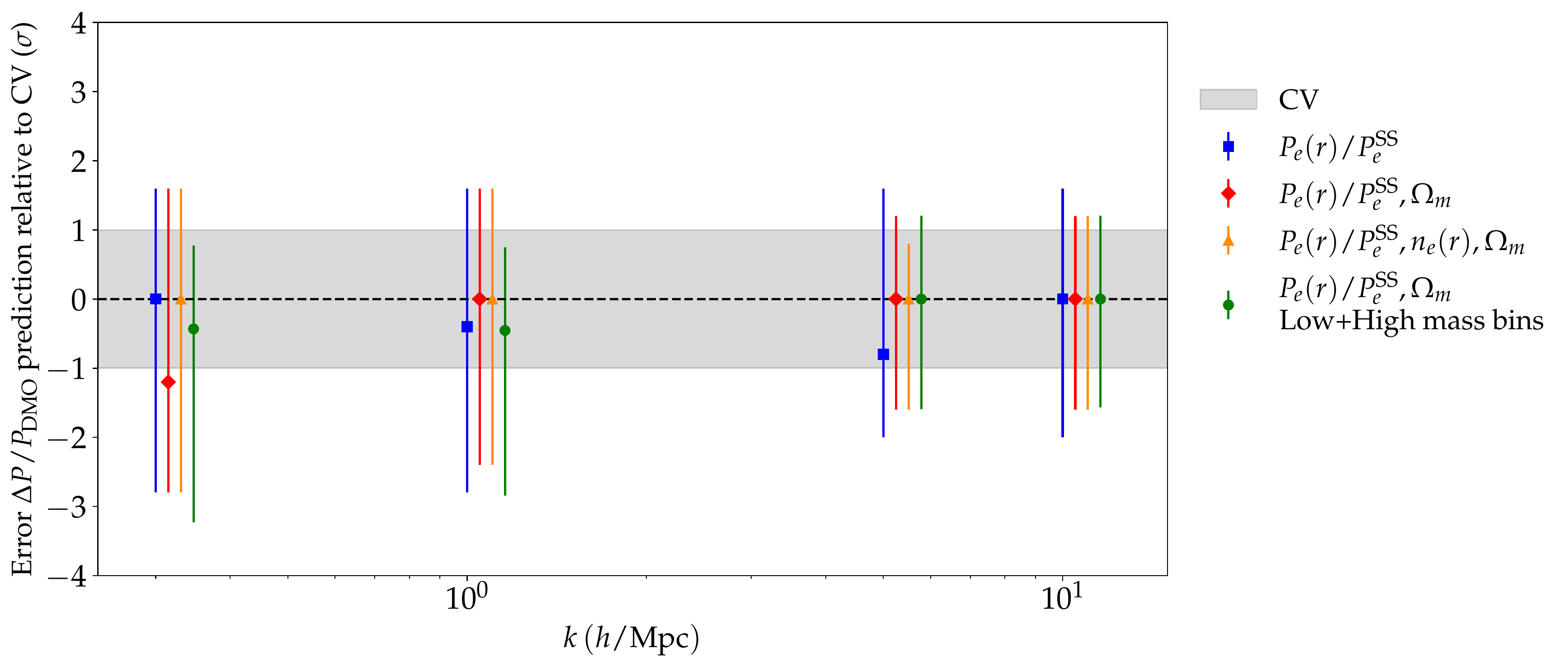}
    \caption{ 
    Same as Fig.~\ref{fig:predict_y500_fb} but showing results from  using the full pressure profile, $P_e(r)$, and electron number density profiles, $n_e(r)$, instead of the integrated quantities. 
    We again find that with pressure profile and $\Omega_{\rm m}$ information we can recover robust and precise constraints on the matter power suppression.
    }
    \label{fig:predict_profiles}
\end{figure*}

Fig.~\ref{fig:Pk_Y_CV} shows the accuracy of the RF predictions  for $\Delta P/P_{\rm DMO}$ when trained on stacked $f_b$ (for halos in $5\times 10^{12} < M_{\rm halo} (M_{\odot}/h) < 10^{14}$) and $\Omega_{\rm m}$, normalized to the sample variance error in $\Delta P/P_{\rm DMO}$. As we will show later in this section, this combination of inputs results in precise constraints on the matter power suppression. Specifically to obtain the constraints, after training the RF regressor on the train simulations, we predict the $\Delta P/P_{\rm DMO}$ on test simulation boxes at four scales. Thereafter, we create a histogram of the difference between truth and predicted $\Delta P/P_{\rm DMO}$, normalized by the variance obtained from the CV set of simulations, for each respective suite of simulations (see Fig.~\ref{fig:Pk_Bk_CV}). In Fig.~\ref{fig:Pk_Y_CV}, each errorbar corresponds to the 16th and 84th percentile from this histogram and the marker corresponds to its peak.  We show the results of training and testing on a single simulation suite, and also the results of training/testing across different simulation suites.  It is clear that when training and testing on the same simulation suite, the RF learns a model that comes close to the best possible uncertainty on $\Delta P/P_{\rm DMO}$ (i.e. cosmic variance).  When training on one or two simulation suites and testing another, however, the predictions show bias at low $k$.  This suggests that the model learned from one simulation does not generalize very well to another in this regime.  This result is somewhat different from the findings of \citet{vanDaalen:2020}, where it was found that the relationship between $f_b$ and $\Delta P/P_{\rm DMO}$ did generalize to different simulations.  This difference may result from the fact that we are considering a wider range of feedback prescriptions than in \citet{vanDaalen:2020}, as well as considering significant variations in cosmological parameters. 

Fig.~\ref{fig:Pk_Y_CV} also shows the results of testing and training on all three simulations (black points with errorbars).  Encouragingly, we find that in this case, the predictions are of comparable accuracy to those obtained from training and predicting on the same simulation suite.  This suggests that there is a general relationship across all feedback models that can be learned to go from $\Omega_{\rm m}$ and $f_b$ to $\Delta P/P_{\rm DMO}$.  Henceforth, we will show results trained on all simulation suites and tested on all simulations suites.  Of course, this result does not imply that our results will generalize to some completely different feedback prescription.

In Fig.~\ref{fig:predict_y500_fb} we show the results of training the random forest on different combinations of $f_b$, $Y_{500c}$ and $\Omega_{\rm m}$.   Consistent with the findings of \citet{vanDaalen:2020}, we find that $f_b/(\Omega_{\rm b}/\Omega_{\rm m})$ results in robust constraints on the matter power suppression (blue points with errors).  These constraints come close to the cosmic variance limit across a wide range of $k$.

We additionally find that providing $f_b$ and $\Omega_{\rm m}$ as separate inputs to the RF  improves the precision of the predictions for $\Delta P/P_{\rm DMO}$ relative to using just the combination $f_b/(\Omega_{\rm b}/\Omega_{\rm m})$, with the largest improvement coming at small scales.  This is not surprising given the predictions of our simple model, for which it is clear that $\Delta P/P_{\rm DMO}$ can be impacted by both $\Omega_{\rm m}$ and $f_b / (\Omega_{\rm b} /\Omega_{\rm b})$ independently.  Similarly, it is clear from Fig.~\ref{fig:Pk_SIMBA_allparams} that changing $\Omega_{\rm m}$ changes the relationship between $\Delta P/P_{\rm DMO}$ and the halo gas-derived quantities (like $Y$ and $f_b$).

We next consider a model trained on $Y_{500c}/Y^{\rm SS}$ (orange points in Fig.~\ref{fig:predict_y500_fb}). This model yields reasonable predictions for $\Delta P/P_{\rm DMO}$, although not quite as good as the model trained on $f_b/(\Omega_{\rm b}/\Omega_{\rm m})$.  The $Y/Y^{\rm SS}$ model yields somewhat larger errorbars, and the distribution of $\Delta P/P_{\rm DMO}$ predictions is highly asymmetric.  When we train the RF model jointly on $Y_{500c}/Y^{\rm SS}$ and $\Omega_{\rm m}$ (green points), we find that the predictions improve considerably, particularly at high $k$.  In this case, the predictions are typically symmetric around the true $\Delta P/P_{\rm DMO}$, have smaller uncertainty compared to the model trained on $f_b/(\Omega_{\rm b}/\Omega_{\rm m})$, and comparable uncertainty to the model trained on $\{ f_b/(\Omega_{\rm b}/\Omega_{\rm m})$,$\Omega_{\rm m} \}$.  We thus conclude that when combined with matter density information, $Y/Y^{\rm SS}$ provides a powerful probe of baryonic effects on the matter power spectrum. 

Above we have considered the integrated tSZ signal from halos, $Y_{500c}$.  Measurements in data, however, can potentially probe the tSZ profiles rather than only the integrated tSZ signal (although the instrumental resolution may limit the extent to which this is possible).  In Fig.~\ref{fig:predict_profiles} we consider RF models trained on the stacking the full electron density and pressure profiles in the halo mass range instead of just the integrated quantities. The electron pressure and number density profiles are measured in eight logarithmically spaced bins between $0.1 < r/r_{200c} < 1$. We find that while the ratio $P_e(r)/P^{\rm SS}$ results in robust predictions for $\Delta P/P_{\rm DMO}$, simultaneously providing $\Omega_{\rm m}$ makes the predictions more precise. Similar to the integrated profile case, we find that additionally providing the electron density profile information only marginally improves the constraints. We also show the results when jointly using the measured pressure profiles for both the low and high mass halos to infer the matter power suppression. We find that this  leads to only a marginal improvements in the constraints. 

Note that we have input the 3D pressure and electron density profiles in this case. Even though observed SZ maps are projected quantities, we can infer the 3D pressure profiles from the model used to analyze the projected correlations.

\begin{figure*}
    \centering
    \includegraphics[width=0.95\textwidth]{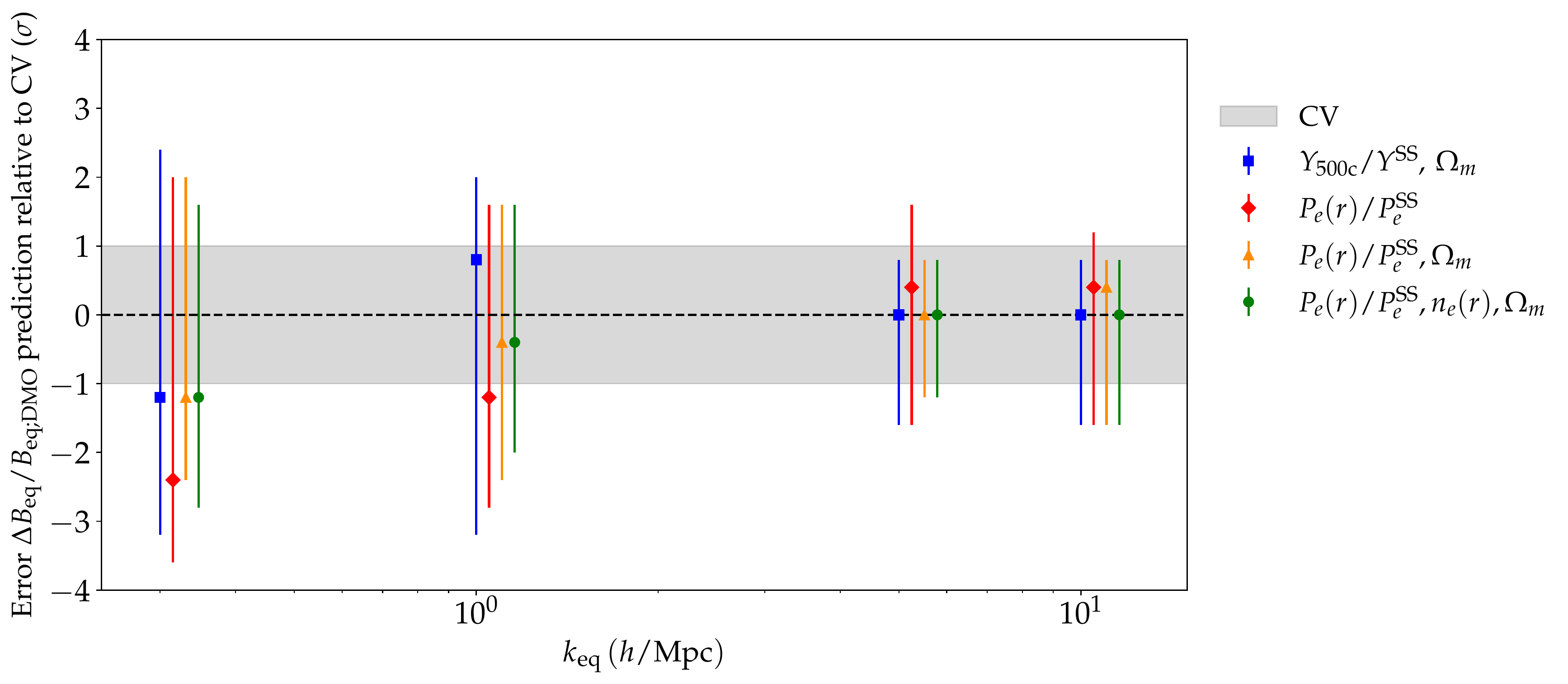}   
    \caption{Same as Fig.~\ref{fig:predict_y500_fb}, but for the impact of feedback on the bispectrum  in equilateral triangle configurations. We find that the inclusion of pressure profile information results in unbiased constraints on feedback effects on the bispectrum. 
    }
    \label{fig:predict_Bk_eq}
\end{figure*}

\subsection{Predicting baryonic effects on the bispectrum with $f_b$ and the electron pressure}
In Fig.~\ref{fig:predict_Bk_eq}, we repeat our 
 analysis from above to make predictions for baryonic effects on the matter bispectrum, $\Delta B(k)/B(k)$. Similar to the matter power spectrum, we train and test our model on a combination of the three simulations. We train and test on equilateral triangle bispectrum configurations with different scales $k$. 
We again see that information about the electron pressure and $\Omega_{\rm m}$ results in precise and unbiased constraints on the impact of baryonic physics on the bispectrum. The constraints improve as we go to the small scales. In Appendix~\ref{app:Bk_sq} we show similar methodology applied to squeezed bispectrum configurations. 

However, there are several important caveats to these results.  The bispectrum is sensitive to high-mass ($M > 5\times 10^{13} M_{\odot}/h$) halos \citep{Foreman:2020:MNRAS:} which are missing from the CAMELS simulations. Consequently, our measurements of baryonic effects on the bispectrum can be biased when using CAMELS.  The simulation resolution can also impact the bispectrum significantly.  A future analysis with larger volume sims at high resolution could use the methodology introduced here to obtain more robust results.  Finally, there would is likely to be covariance between the power spectrum suppression and baryonic effects on the bispectrum, as they both stem from same underlying physics.  We defer a complete exploration of these effects to future work.  

\section{Results II: ACTxDES measurements and forecast}
\label{sec:results_data}

\begin{figure*}
\includegraphics[scale = 0.45]{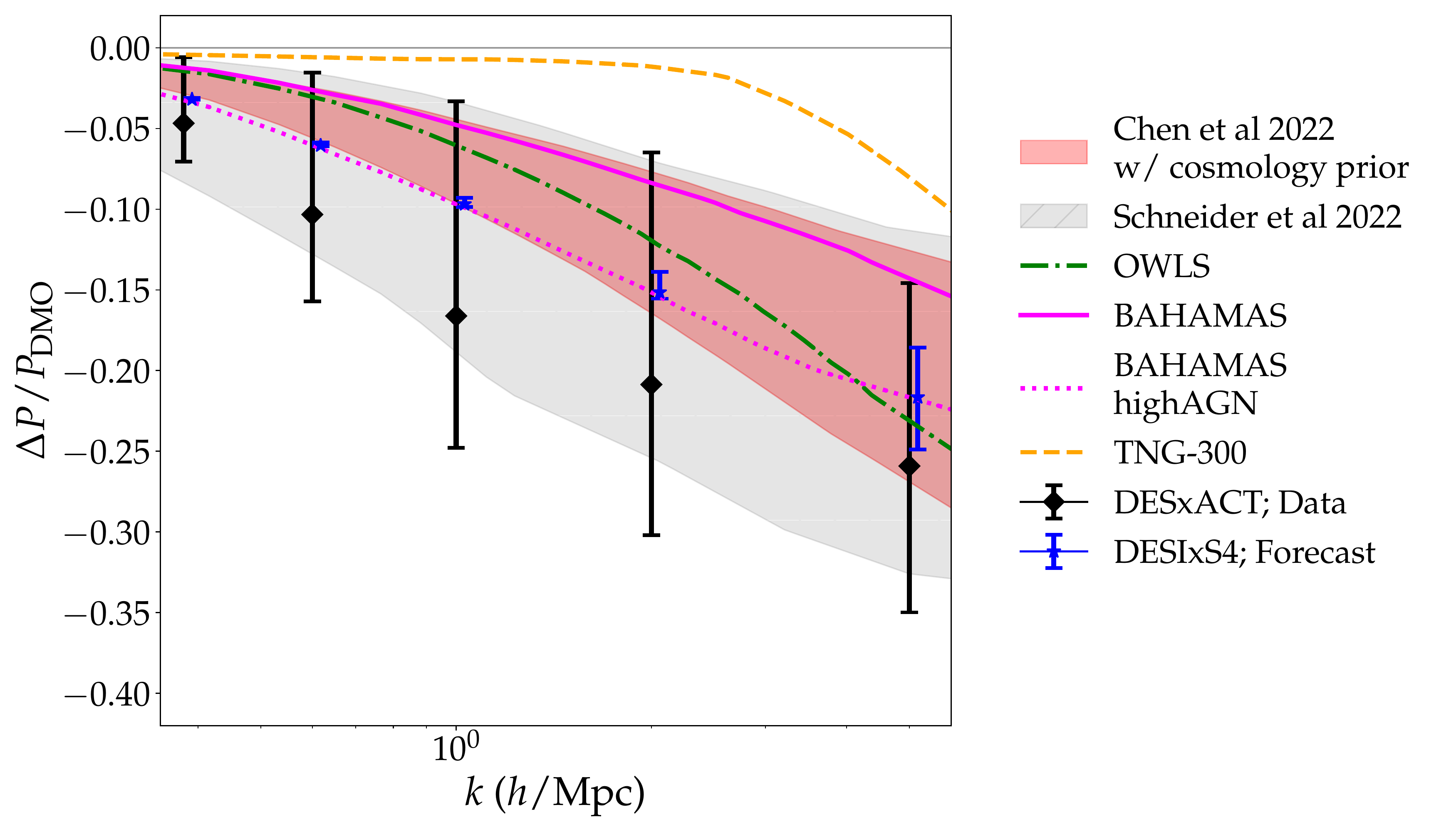}
\caption[]{Constraints on the impact of feedback on the matter power spectrum obtained using our trained random forest model applied to measurements of $Y_{\rm 500c}/Y^{\rm SS}$ from the DESxACT analysis of \citet{Pandey:2022} (black points with errorbars). We also show the expected improvements from future halo-$y$ correlations from DESIxSO using the constraints in \citet{Pandey:2020}. We compare these to the inferred constraints obtained using cosmic shear \citep{Chen:2023:MNRAS:} and additionally including X-ray and kSZ data \citep{Schneider:2022:MNRAS:}. We also compare with the results from larger simulations: OWLS \citep{Schaye:2010:MNRAS:}, BAHAMAS \citep{McCarthy:2017:MNRAS:} and TNG-300 \citep{Springel:2018:MNRAS:}.
}
\label{fig:Pk_data_forecast}
\end{figure*}

Our analysis above has resulted in a statistical model (i.e. a random forest regressor) that predicts the matter power suppression $\Delta P/P_{\rm DMO}$ 
given values of $Y_{500c}$ for low-mass halos.  This model is robust to significant variations in the feedback prescription, at least across the SIMBA, TNG and Astrid models.  We now apply this model to constraints on $Y_{500c}$ coming from the cross-correlation of galaxy lensing shear with tSZ maps measured using Dark Energy Survey (DES) and Atacama Cosmology Telescope (ACT) data.

\citet{Gatti:2022} and \citet{Pandey:2022} measured 
the cross-correlations of DES galaxy lensing with Compton $y$ maps from a combination of Advanced ACT \citep{Madhavacheril:2020:PhRvD:} and {\it Planck} data \citep{PlanckCollaboration:2016:A&A:} over an area of 400 sq. deg. They analyze these cross-correlations using a halo model framework, where the pressure profile in halos was parameterized using a generalized Navarro-Frenk-White profile \citep{Navarro:1996:ApJ:, Battaglia:2012:ApJ:b}. This pressure profile is described using four free parameters, allowing for scaling with mass, redshift and distance from halo center.  The constraints on the parameterized pressure profiles can be translated directly into constraints on $Y_{500c}$ for halos in the mass range relevant to our random forest models.  

We use the parameter constraints from \citet{Pandey:2022} to generate 400 samples of the inferred 3D profiles of halos at $z=0$ (i.e. the redshift at which the RF models are trained) in ten logarithmically-spaced mass bins in range $12.7 < \log_{10}(M/h^{-1} M_{\odot}) < 14$.  We then perform the volume integral of these profiles to infer the $Y_{\rm 500c}(M, z)$ (see Eq.~\ref{eq:Y500_from_Pe}).
Next, we generate a halo-averaged value of $Y_{500c}/Y^{\rm SS}$ for the $j$th sample by integrating over the halo mass distribution in CAMELS:
\begin{equation}\label{eq:Pe_stacked_data}
    \bigg\langle \frac{Y_{\rm 500c}}{Y^{\rm SS}} \bigg\rangle^j = \frac{1}{\bar{n}^j} \int dM \bigg(\frac{dn}{dM}\bigg)^j_{\rm CAMELS} \frac{Y^j_{\rm 500c}(M)}{Y^{\rm SS}}
\end{equation}
where $\bar{n}^j = \int dM (dn/dM)^j_{\rm CAMELS}$ and $(dn/dM)^j_{\rm CAMELS}$ are a randomly chosen halo mass function from the CV set of boxes of TNG, SIMBA or Astrid. This procedure allows us to incorporate the impact and uncertainties of the CAMELS box size on the halo mass function. Note that due to the small box size of CAMELS, there is a deficit of high mass halos and hence the functional form of the mass function differs somewhat from other fitting functions in literature, e.g. \cite{Tinker:2008:ApJ:}. 

Fig.~\ref{fig:Pk_data_forecast} shows the results feeding the $Y_{500c}/Y^{\rm SS}$ values calculated above into our trained RF model to infer the impact of baryonic feedback on the matter power spectrum (black points with errorbars).  The RF model used is that trained on the TNG, SIMBA and Astrid simulations. The errorbars represent the 16th and 84th percentile of the recovered $\Delta P/P_{\rm DMO}$ distribution using the 400 samples described above. Note that in this inference we fix the matter density parameter, $\Omega_{\rm m} = 0.3$, same value as used by the CAMELS CV simulations as we use these to estimate the halo mass function.  

In the same figure, we also show the constraints from \citet{Chen:2023:MNRAS:} and \citet{Schneider:2022:MNRAS:} obtained using the analysis of complementary datasets. \citet{Chen:2023:MNRAS:} analyze the small scale cosmic shear measurements from DES Year-3 data release using a baryon correction model. Note that in this analysis, they only use a limited range of cosmologies, particularly restricting to high $\sigma_8$ due to the requirements of emulator calibration. Moreover they also impose cosmology constraints from the large scale analysis of the DES data. 
Note that unlike the procedure presented here, their modeling and constraints are sensitive to the priors on $\sigma_8$.  
\citet{Schneider:2022:MNRAS:} analyze the X-ray data (as presented in \citealt{Giri:2021:JCAP:}) and kSZ data from ACT and SDSS \citep{Schaan:2021:PhRvD:} and the cosmic shear measurement from KiDS \citep{Asgari:2021}, using another version of baryon correction model. A joint analysis from these complementary dataset leads to crucial degeneracy breaking in the parameters. It would be interesting to include the tSZ observations presented here in the same framework as it can potentially make the constraints more precise.

Several caveats about our analysis with data are in order.  First, the lensing-SZ correlation is most sensitive to halos in the mass range of $M_{\rm halo} \geq 10^{13} M_{\odot}/h$.  However, our RF model operates on halos with mass in the range of $5 \times 10^{12} \geq M_{\rm halo} \leq 10^{14} M_{\odot}/h$, with the limited volume of the simulations restricting the number of halos above $10^{13} M_{\odot}/h$. We have attempted to account for this selection effect by using the halo mass function from the CV sims of the CAMELS simulations when calculating the stacked profile. However, using a larger volume simulation suite would be a better alternative (also see discussion in Appendix~\ref{app:volume_res_comp}). Moreover, the CAMELS simulation suite also fix the value of $\Omega_{\rm b}$. There may be a non-trivial impact on the inference of $\Delta P/P_{\rm DMO}$ when varying that parameter.  Note, though, that $\Omega_b$ is tightly constrained by other cosmological observations. Lastly, the sensitivity of the lensing-SZ correlations using DES galaxies is between $0.1 < z < 0.6$. However, in this study we extrapolate those constraints to $z=0$ using the pressure profile model of \citet{Battaglia:2012:ApJ:b}. We note that inference obtained at the peak sensitivity redshift would be a better alternative but we do not expect this to have a significant impact on the conclusions here.

In order to shift the sensitivity of the data correlations to lower halo masses, it would be preferable to analyze the galaxy-SZ and halo-SZ correlations. In \citet{Pandey:2020} we forecast the constraints on the inferred 3D pressure profile from the future halo-SZ correlations using DESI and CMB-S4 SZ maps for a wide range of halo masses. In Fig.~\ref{fig:Pk_data_forecast} we also show the expected constraints on the matter power suppression using the halo-SZ correlations from halos in the range $M_{500c} > 5\times 10^{12} M_{\odot}/h$. We again follow the same methodology as described above to create a stacked normalized integrated pressure (see Eq.~\ref{eq:Pe_stacked_data}). Moreover, we also fix $\Omega=0.3$ to predict the matter power suppression. Note that we shift the mean value of $\Delta P/P_{\rm DMO}$ to the recovered value from BAHAMAS high-AGN simulations \citep{McCarthy:2017:MNRAS:}. 
As we can see in Fig.~\ref{fig:Pk_data_forecast}, we can expect to obtain significantly more precise constraints from these future observations. 

\section{Conclusions}
\label{sec:conclusion}

We have shown that the tSZ signals from low-mass halos contain significant information about the impacts of baryonic feedback on the small-scale matter distribution.  Using models trained on hydrodynamical simulations with a wide range of feedback implementations, we demonstrate that information about baryonic effects on the power spectrum and bispectrum can be robustly extracted.  By applying these same models to measurements with ACT and DES, we have shown that current tSZ measurements already constrain the impact of feedback on the matter distribution.  Our results suggest that using simulations to learn the relationship between halo gas observables and baryonic effects on the matter distribution is a promising way forward for constraining these effects with data.

Our main findings from our explorations with the CAMELS simulations are the following:

\begin{itemize}
    \item In agreement with \citet{vanDaalen:2020}, we find that baryon fraction in halos correlates with the power spectrum suppression. We find that the correlation is especially robust at small scales.
    
    \item We find (in agreement with \citealt{Delgado:23}) that there can be significant scatter in the relationship between baryon fraction and power spectrum suppression at low halo mass, and that the relationship varies to some degree with feedback implementation.  However, the bulk trends appear to be consistent regardless of feedback implementation.
    
    \item We propose a simple model that qualitatively (and in some cases quantitatively) captures the broad features in the relationships between the impact of feedback on the power spectrum, $\Delta P/P_{\rm DMO}$, at different values of $k$, and halo gas-related observables like $f_b$ and $Y_{500c}$ at different halo masses. 
    \item Despite significant scatter in the relations between $Y_{500c}$ and $\Delta P/P_{\rm DMO}$ at low halo mass, we find that simple random forest models yield tight and robust constraints on $\Delta P/P_{\rm DMO}$ given information about $Y_{500c}$ in low-mass halos and $\Omega_{\rm m}$.  
    \item Using the pressure profile instead of just the integrated $Y_{\rm 500c}$ signal provides additional information about $\Delta P/P_{\rm DMO}$, leading to 20-50\% improvements when not using any cosmological information. When additionally providing the $\Omega_{\rm m}$ information, the improvements in constraints on baryonic changes to the power spectrum or bispectrum are modest when using the full pressure profile relative to integrated quantities like $Y_{500c}$.  
    \item The pressure profiles and baryon fractions also carry information about baryonic effects on the bispectrum.  
\end{itemize}

Our main results from our analysis of constraints from the DESxACT shear-$y$ correlation analysis are
\begin{itemize}
    \item We have used the DES-ACT measurement of the  shear-tSZ correlation from \cite{Gatti:2022} and \cite{Pandey:2022} to infer $Y_{500c}$ for halos in the mass range relevant to our random forest models.  Feeding the measured $Y_{500c}$ into these models, we have inferred the impact of baryonic effects on the power spectrum, as shown in Fig.~\ref{fig:Pk_data_forecast}.
    \item We show that constraints on baryonic effects on the power spectrum will improve significantly in the future, particularly using halo catalogs from DESI and tSZ maps from CMB-S4. 
\end{itemize}

With data from future galaxy and CMB surveys, we expect constraints on the tSZ signal from halos across a wide mass and redshift range to improve significantly.  These improvements will come from both the galaxy side (e.g. halos detected over larger areas of the sky, down to lower halo masses, and out to higher redshifts) and the CMB side (more sensitive tSZ maps over larger areas of the sky).  Our forecast for DESI and CMB Stage 4 in Fig.~\ref{fig:Pk_data_forecast} suggests that very tight constraints can be obtained on the impact of baryonic feedback on the matter power spectrum.  We expect that these constraints on the impact of baryonic feedback will enable the extraction of more cosmological information from the small-scale matter distribution.

\section{Acknowledgements}
DAA acknowledges support by NSF grants AST-2009687 and AST-2108944, CXO grant TM2-23006X, and Simons Foundation award CCA-1018464.

\section{Data Availability}
The TNG and SIMBA simulations used in this work are part of the CAMELS public data release \citep{Villaescusa-Navarro:2021:ApJ:} and are available at \url{https://camels.readthedocs.io/en/latest/}. The Astrid simulations used in this work will be made public before the end of the year 2023.  The data used to make the plots presented in this paper are available upon request.

\bibliographystyle{mnras}
\bibliography{thebib,ads}

\appendix

\section{Impact of limited volume of CAMELS simulations}
\label{app:volume_res_comp}

\begin{figure}
\includegraphics[width=0.95\columnwidth]{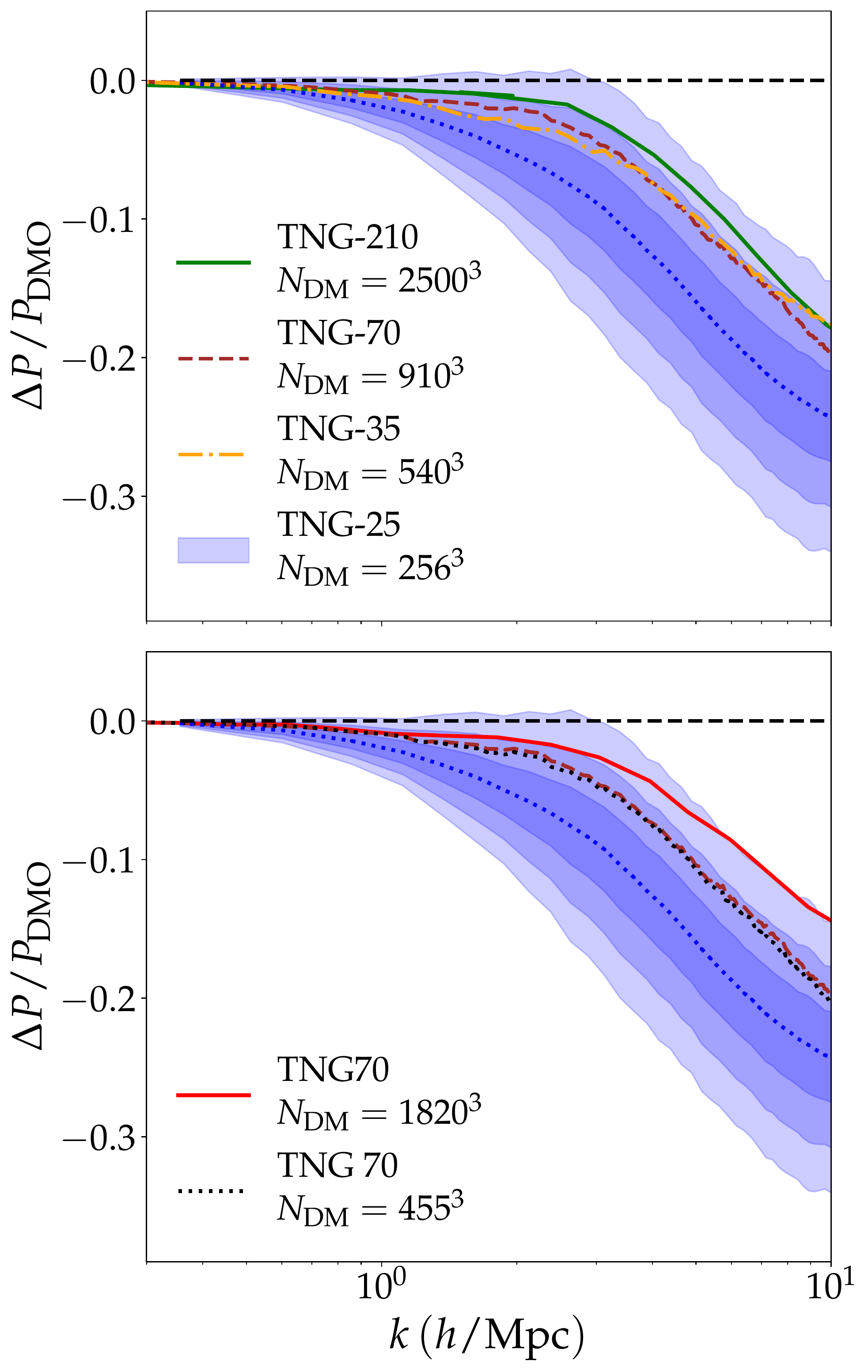}
\caption[]{Comparison of the suppression of matter power in the CAMELS TNG simulation and simulations using the same sub-grid prescription but larger box sizes \citep{Springel:2018:MNRAS:}. We also show 1$\sigma$ and 2$\sigma$ uncertainty due to cosmic variance. In the top panel we change the TNG box sizes, while preserving the resolution, where as in the bottom panel we preserve the TNG box size while changing the resolution. 
}
\label{fig:Pk_TNG_boxsize}
\end{figure}

In order to analyze the impact of varying box sizes and resolution on the matter power suppression, we use the TNG simulations as presented in \citet{Springel:2018:MNRAS:}. Particularly we use their boxes with side lengths of 210~Mpc/$h$, 75~Mpc/$h$ and 35~Mpc/$h$  (which they refer to as TNG-300, TNG-100 and TNG-50 as it corresponds to side length in the units of Mpc). We then make the comparison to 25~Mpc/$h$ TNG boxes run from CAMELS. We use the CV set of simulations and use them to infer the expected variance due to stochasticity induced by changing initial conditions. Note that the hydrodynamical model is identical between CAMELS CV runs and the bigger TNG boxes. In Fig.~\ref{fig:Pk_TNG_boxsize}, we show the power suppression for these boxes, including the runs at varying resolution. We find that while changing box sizes gives relatively robust values of power suppression, changing resolution can have non-negligible impact. However, all the TNG boxes are consistent at 2-3$\sigma$ level relative to the CAMELS boxes. 

\section{Example of Emulation}
\label{app:emulation}
\begin{figure}
    \centering
    \includegraphics[width=0.95\columnwidth]{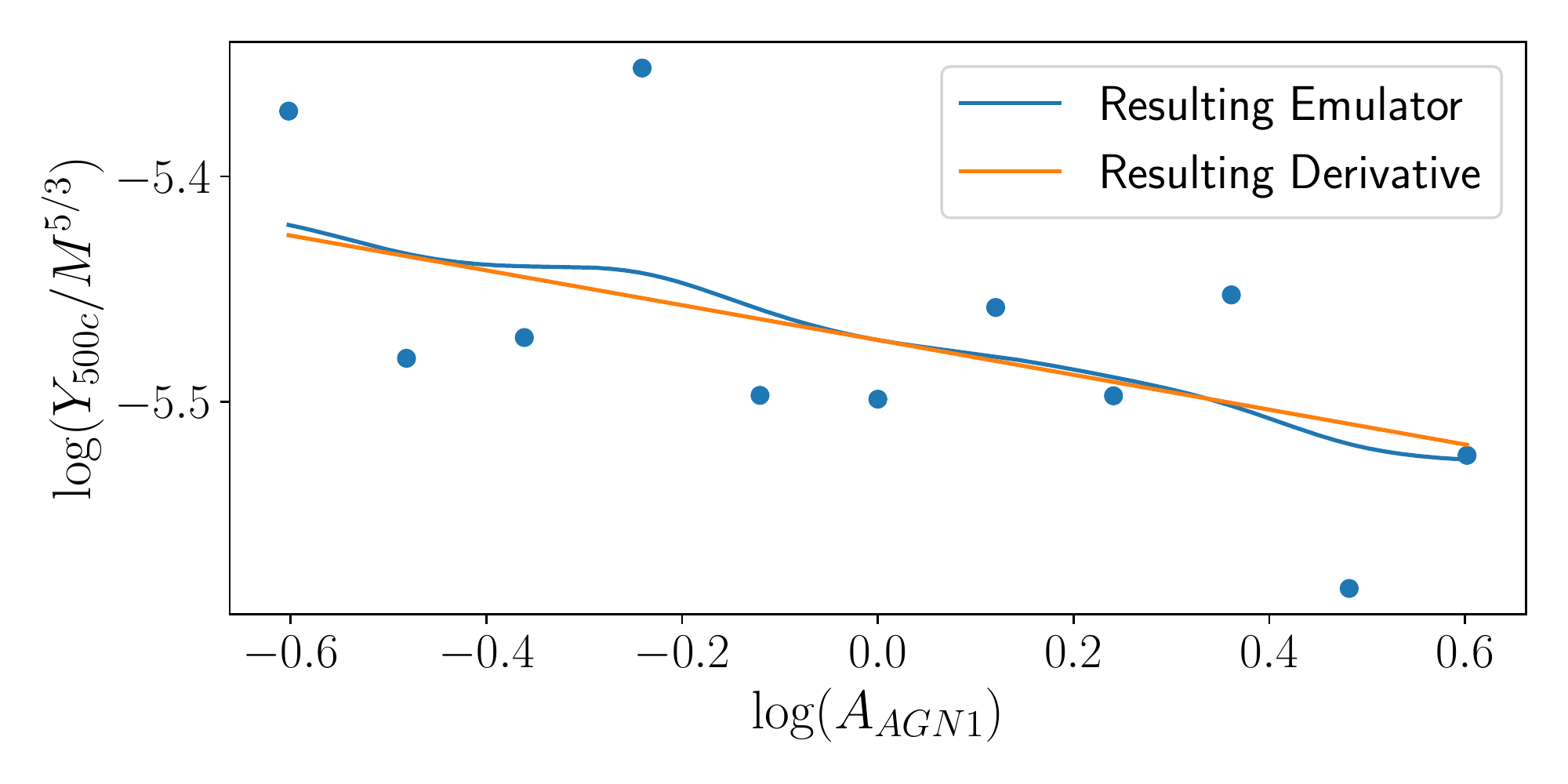}
    \caption{The constructed emulator and resulting derivative for the $A_{\rm AGN1}$ parameter in the mass bin $10^{12}<M(M_\odot /h)<5\times 10^{12}$. }
    \label{fig:emu}
\end{figure}

We present an example of the constructed emulator from \S\ref{sec:fisher} for the $A_{\rm AGN1}$ parameter in Fig. \ref{fig:emu}. This shows how we estimate the derivative of the observable ($Y_{500c}/M^{5/3}$) in a way that is robust to  stochasticity. 

\section{Robustness of results to different train simulations}
\begin{figure}
\includegraphics[width=0.95\columnwidth]{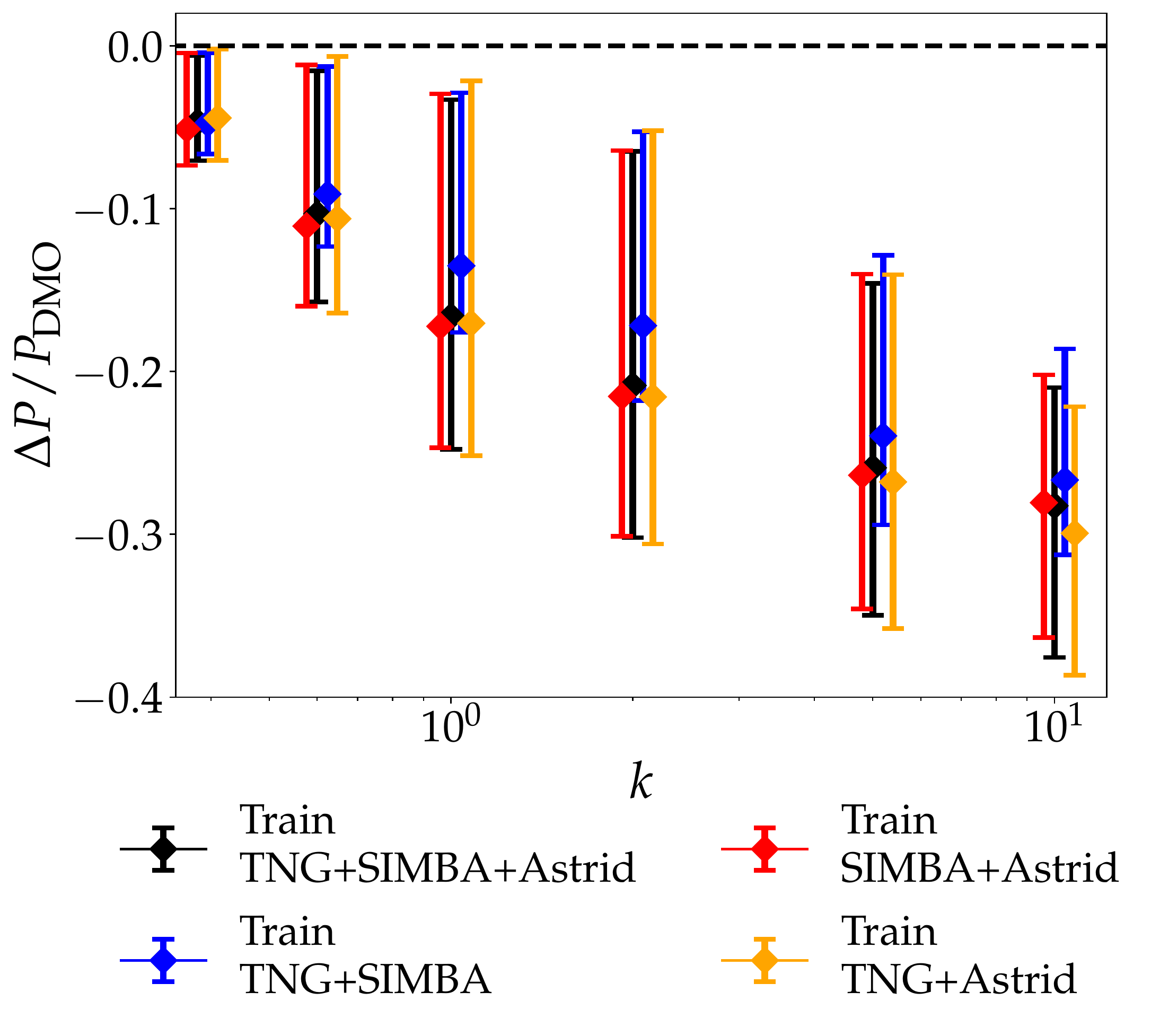}
\caption[]{In this figure we change the simulations used to train the RF when inferring the power suppression from the data measurements.
}
\label{fig:Pk_data_change_sims}
\end{figure}

In Fig.~\ref{fig:Pk_data_change_sims}, we test the impact of changing the simulations used to train the random forest regressor. We then use these different trained models to infer the constraints on the matter power suppression from the same stacked $\langle Y_{500c}/Y^{\rm SS} \rangle$ as described in \S~\ref{sec:results_data}. We see that our inferred constraints remain consistent when changing the simulations. 

\section{Test with lower halo masses}
\label{app:low_mass}
\begin{figure*}
    \centering
    \includegraphics[width=0.95\textwidth]{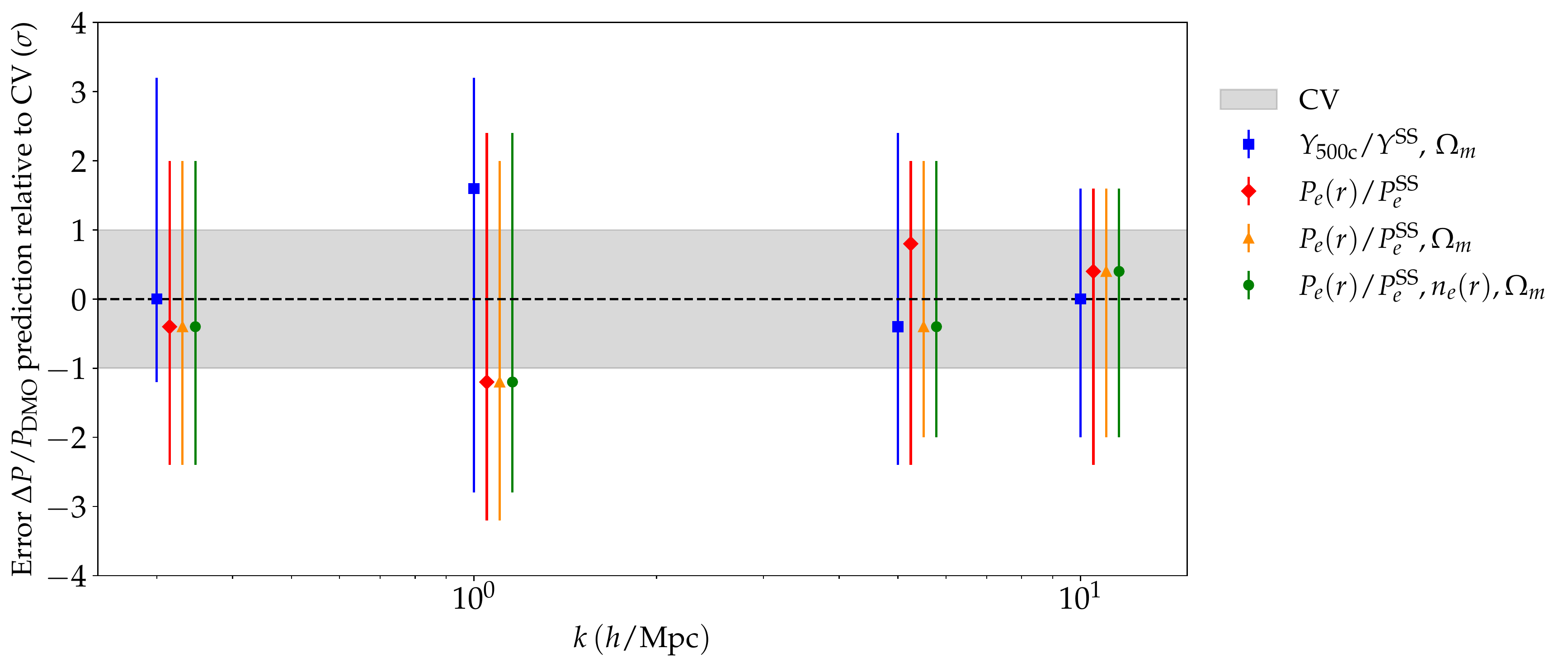}   
    \caption{Same as Fig.~\ref{fig:predict_y500_fb} and Fig.~\ref{fig:predict_profiles}, but obtained on lower halo masses, $1\times10^{12} < M (M_{\odot}/h) < 5\times 10^{12}$. We find that having pressure profile information results in unbiased constraints here as well, albeit with a larger errorbars}
    \label{fig:predict_lowmass}
\end{figure*}

In Fig.~\ref{fig:predict_lowmass}, we show the constraints on the power suppression obtained by analyzing the observables obtained from halos with lower masses, $1\times10^{12} < M (M_{\odot}/h) < 5\times 10^{12}$. We see that remarkably, even these lower halo masses provide unbiased constraints on the matter power suppression with robust inference especially at smaller scales. However, when compared to the results descibed in \S~\ref{sec:fbY}, we obtain less precise constraints. This is expected as lower halos with lower masses are more susceptible to environmental effects which induces a larger scatter in the relation between their observables (such as $f_b$ or $Y_{500c}$) and their halo masses governing feedback processes.

\section{Test with other bispectrum configurations}\label{app:Bk_sq}
\begin{figure*}
    \centering
    \includegraphics[width=0.95\textwidth]{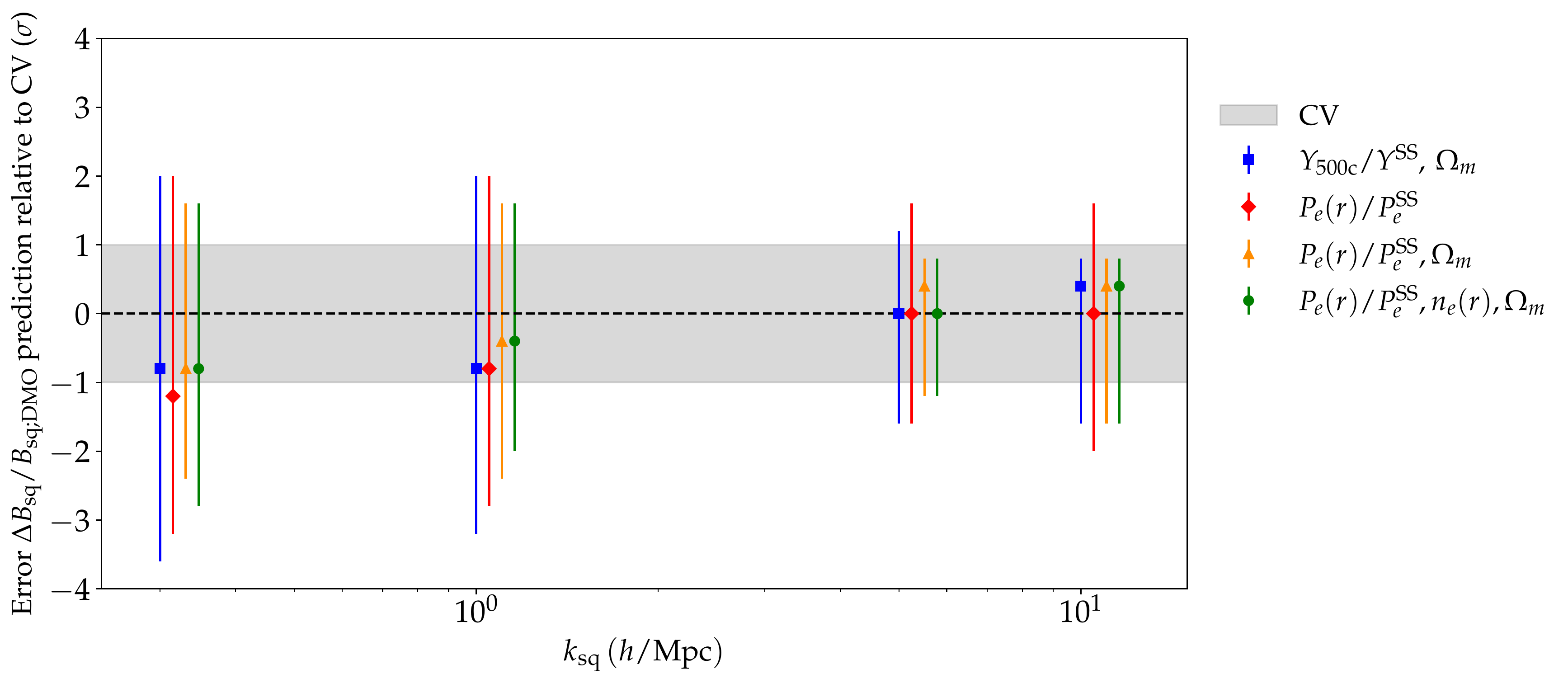}   
    \caption{Same as Fig.~\ref{fig:predict_Bk_eq}, but for squeezed triangle configurations ($\mu = 0.9$).}
    \label{fig:predict_Bk_sq}
\end{figure*}

In Fig.~\ref{fig:predict_Bk_sq}, we show the constraints obtained on the suppression of the squeezed bispectrum configurations. We fix the the angle between the long sides of the triangle to correspond to $\mu = 0.9$. We again find robust inference of baryonic effects on the bispectrum  when using either the integrated pressure profile or full radial pressure profile. 

\bsp    % typesetting comment
\label{lastpage}
\end{document}